\documentclass[twocolumn]{revtex4-1}
\usepackage{graphicx}
\usepackage{url}  
\usepackage{amsmath}
\usepackage{amssymb}
\usepackage{subfigure}
\newcommand\scalemath[2]{\scalebox{#1}{\mbox{\ensuremath{\displaystyle #2}}}}

\begin{document}
\title{Emergent behaviour in a chlorophenol-mineralising three-tiered microbial `food web'}

\author{M. J. Wade}
\email[]{matthew.wade@ncl.ac.uk}
\affiliation{School of Civil Engineering of Geosciences, Newcastle University, Newcastle-upon-Tyne NE1 7RU, United Kingdom}
\author{R. W. Pattinson}
\affiliation{School of Civil Engineering of Geosciences, Newcastle University, Newcastle-upon-Tyne NE1 7RU, United Kingdom}
\affiliation{School of Mathematics and Statistics, Newcastle University, Newcastle-upon-Tyne NE1 7RU, United Kingdom}
\author{N. G. Parker}
\affiliation{School of Mathematics and Statistics, Newcastle University, Newcastle-upon-Tyne NE1 7RU, United Kingdom}
\author{J. Dolfing}
\affiliation{School of Civil Engineering of Geosciences, Newcastle University, Newcastle-upon-Tyne NE1 7RU, United Kingdom}

\begin{abstract}
Anaerobic digestion enables the water industry to treat wastewater as a resource for generating energy and recovering valuable by-products. The complexity of the anaerobic digestion process has motivated the development of complex models.  However, this complexity makes it intractable to pin-point stability and emergent behaviour.  Here, the widely used Anaerobic Digestion Model No. 1 (ADM1) has been reduced to its very backbone, a syntrophic two-tiered microbial `\textit{food chain}' and a slightly more complex three-tiered microbial `\textit{food web}', with their stability analysed as function of the inflowing substrate concentration and dilution rate. Parameterised for phenol and chlorophenol degradation, steady-states were always stable and non-oscillatory.  Low input concentrations of chlorophenol were sufficient to maintain chlorophenol- and phenol-degrading populations but resulted in poor conversion and a hydrogen flux that was too low to sustain hydrogenotrophic methanogens. The addition of hydrogen and phenol boosted the populations of all three organisms, resulting in the counterintuitive phenomena that (i) the phenol degraders were stimulated by adding hydrogen, even though hydrogen inhibits phenol degradation, and (ii) the dechlorinators indirectly benefitted from measures that stimulated their hydrogenotrophic competitors; both phenomena hint at emergent behaviour. 
\end{abstract}
\maketitle

\medmuskip=0mu
\arraycolsep=1pt

\section{Introduction}\label{Intro}
Microbial degradation of organic compounds in methanogenic environments is a sequential process catalysed by a series of different micro-organisms. Syntrophy plays a pivotal role in these feeding webs: degradation of compounds like propionate and phenol is only sustainable if their degradation products, hydrogen and acetate, are removed by methanogens. The thermodynamic rational behind syntrophy is well understood, but its kinetic framework is less established. This raises questions about the stability of these feeding chains and the factors that govern them. As a first step towards answering these questions a simple mathematical model was previously developed describing the interactions in a two-tiered feeding chain, populated with a set of parameters that apply to propionate degraders and hydrogenotrophic methanogens~\cite{xu11}. Mathematical analysis of the model indicated that the system was always stable: there were no conditions where the populations of the two organisms oscillate or show other forms of emergent behaviour.

The objective of the present paper is to introduce an additional organism into a similar feeding chain and evaluate its effect on stability and potential emergent behaviour of the resulting `food web'. The organism of choice is a chlorophenol-dechlorinating bacterium. The other two organisms are a phenol degrader and a hydrogenotrophic methanogen. The complete removal of phenolic compounds from the system is hereby referred to as chlorophenol mineralisation. The salient feature of the chlorophenol degrader here is that production of phenol is coupled to consumption of hydrogen by hydrogen cycling. Thus, as a hydrogen consumer, the dechlorinator competes with the methanogen for hydrogen~\cite{dolfing86,dolfing91}. The working hypotheses are (i) that the dechlorinator can (partially) replace the methanogen as the syntroph in a phenol-degrading consortium, and (ii) that introduction of this organism can potentially lead to unexpected emergent behaviour related to the intricacies of the multi-species relationships. 

It has been shown that deterministic modelling of biological systems, typically through a system of coupled ordinary differential equations (ODEs), provides important understanding of these often complex processes, specifically in determining changes to the system behaviour given perturbations in the inputs. For anaerobic digestion, higher dimensional models are useful for capturing the phenomenological behaviour of the multi-step processes and are often the de-facto method for understanding plant operation~\cite{batstone02}. On the other hand, simplified or reduced models have received more attention in process monitoring, control design~\cite{bernard01,gaida12}, and optimisation~\cite{vavilin01}. Simplified models have also been applied to determine the global~\cite{benyahia12, shen07} or local~\cite{simeonov10} stability of the system under investigation. Typically, these are two-species models, whereas here we present a case considering a three-species food web.

Whilst a strictly analytical approach is not possible given the dimensions of the ensuing model, the approach taken here derives analytic expressions for all steady-states, supported by numerical simulations to determine the regions of local stability within sensible operating conditions. As such, it is possible to gain a formal understanding of the emergent properties of all states.

\section{Model and Method}\label{Models}

\subsection{Mechanistic model of a three-tiered food web}

The model developed here is based on Anaerobic Digestion Model No. 1 (ADM1)~\cite{batstone02}. The general model here has six components, three substrate and three biomass variables, from which a sub-model describing phenol degradation and the extension of the full model to include addition of extraneous substrates, are formed and described in the relevant sections. The chlorophenol degrader utilises both chlorophenol and hydrogen for growth, producing phenol as a product. Phenol is consumed by the phenol degrader forming hydrogen, which also is inhibitory to its growth. The methanogen scavenges this hydrogen and acts as the primary syntroph. 

The time-dependent substrate and biomass concentrations are denoted $S(t)$ and $X(t)$, respectively, with subscripts $_\mathrm{ch}$, $_\mathrm{ph}$ and $_\mathrm{H_2}$ referring to chlorophenol, phenol and hydrogen components, respectively.  The growth functions are of Monod form with the inclusion of a product inhibition term, $K_{I,\mathrm{H_{2}}}$.  Chlorophenol, phenol and hydrogen are introduced with an input concentration $S_{\mathrm{ch,in}}$, $S_{\mathrm{ph,in}}$, $S_{\mathrm{H_2,in}}$, respectively, and a dilution rate $D$.  The inhibition of hydrogen on the phenol degrader is defined as

\begin{align}
I_{2} & = \frac{1}{1 + \dfrac{S_{\mathrm{H_{2}}}}{K_{I,\mathrm{H_{2}}}}}.
\end{align}
\noindent

The substrate and biomass concentrations then evolve according to the six-dimensional dynamical system of ODEs

\begin{align}
\label{eqm2} \frac{\mathrm{d}X_{\mathrm{ch}}}{\mathrm{d}t} = & -DX_{\mathrm{ch}} + Y_{\mathrm{ch}}f_0X_{\mathrm{ch}} - k_{\mathrm{dec,ch}}X_{\mathrm{ch}} \\
\label{eqm4} \frac{\mathrm{d}X_{\mathrm{ph}}}{\mathrm{d}t} = & -DX_{\mathrm{ph}} + Y_{\mathrm{ph}}f_1X_{\mathrm{ph}} - k_{\mathrm{dec, ph}}X_{\mathrm{ph}} \\
\label{eqm6} \frac{\mathrm{d}X_{\mathrm{H_{2}}}}{\mathrm{d}t} = & -DX_{\mathrm{H_{2}}} + Y_{\mathrm{H_{2}}}f_2X_{\mathrm{H_{2}}} - k_{\mathrm{dec,H_{2}}}X_{\mathrm{H_{2}}}\\
\label{eqm1} \frac{\mathrm{d}S_{\mathrm{ch}}}{\mathrm{d}t} = & D(S_{\mathrm{ch,in}} - S_{\mathrm{ch}}) - f_0X_{\mathrm{ch}} \\
\label{eqm3} \frac{\mathrm{d}S_{\mathrm{ph}}}{\mathrm{d}t} = & D(S_{\mathrm{ph,in}} - S_{\mathrm{ph}}) + \frac{224}{208}(1-Y_{\mathrm{ch}})f_0X_{\mathrm{ch}} \nonumber \\ & - f_1X_{\mathrm{ph}} \\
\label{eqm5} \frac{\mathrm{d}S_{\mathrm{H_{2}}}}{\mathrm{d}t} = & D(S_{\mathrm{H_{2},in}} - S_{\mathrm{H_{2}}}) + \frac{32}{224}(1 - Y_{\mathrm{ph}})f_1X_{\mathrm{ph}} \nonumber \\ & - f_2X_{\mathrm{H_{2}}} - \frac{16}{208}f_0X_{\mathrm{ch}}. 
\end{align} 

\noindent
Here the functions $f_0$, $f_1$ and $f_2$ are defined as

\begin{align}
\label{eqf1} f_0(S_{\mathrm{ch}},S_{\mathrm{H_2}}) & = \frac{k_{m,\mathrm{ch}}S_{\mathrm{H_{2}}}}{K_{S,\mathrm{H_{2}},c}+S_{\mathrm{H_{2}}}}\frac{S_{\mathrm{ch}}}{K_{S,\mathrm{ch}} + S_{\mathrm{ch}}} \\
\label{eqf2} f_1(S_{\mathrm{ph}},S_{\mathrm{H_2}}) & = \frac{k_{m,\mathrm{ph}}S_{\mathrm{ph}}}{K_{S,\mathrm{ph}} + S_{\mathrm{ph}}}I_{2} \\
\label{eqf3} f_2(S_{\mathrm{H_2}}) & = \frac{k_{m,\mathrm{H_{2}}}S_{\mathrm{H_{2}}}}{K_{S,\mathrm{H_{2}}} + S_{\mathrm{H_{2}}}},
\end{align}
\noindent
where $K_{S,\mathrm{H_{2}},c}$ and $K_{S,\mathrm{H_{2}}}$ are the half-saturation constants for hydrogen in the chlorophenol degrader and hydrogenotrophic methanogen, respectively. The values used for the various parameters are listed in Table~\ref{params}, and their derivation is presented~\ref{appendix:ParamEst}. The value $224/208$ represents the fraction of chlorophenol chemical oxygen demand (COD) converted to phenol, $32/224$ is the fraction of phenol converted to hydrogen, and $16/208$ is the fraction of hydrogen COD consumed by $X_{\mathrm{ch}}$.

\begin{table}
\centering
\begin{tabular}{lll}
\hline
Parameters & Nominal values     & Units                \\ \hline
$k_{m,\mathrm{ch}}$          & 29                         & $\mathrm{kgCOD_{S}/kgCOD_{X}/d}$                     \\
$K_{S,\mathrm{ch}}$          & 0.053                       & $\mathrm{kgCOD/m^{3}}$ \\
$Y_{\mathrm{ch}}$            & 0.019                       & $\mathrm{kgCOD_X/kgCOD_S}$                       \\
$k_{m,\mathrm{ph}}$          & 26                          & $\mathrm{kgCOD_{S}/kgCOD_{X}/d}$                     \\
$K_{S,\mathrm{ph}}$          & 0.302                        & $\mathrm{kgCOD/m^{3}}$ \\
$Y_{\mathrm{ph}}$            & 0.04                        & $\mathrm{kgCOD_X/kgCOD_S}$                       \\
$k_{m,\mathrm{H_{2}}}$    & 35                          & $\mathrm{kgCOD_{S}/kgCOD_{X}/d}$                     \\
$K_{S,\mathrm{H_{2}}}$       & 2.5$\times10^{-5}$ & $\mathrm{kgCOD/m^{3}}$ \\
$K_{S,\mathrm{H_{2},c}}$  & 1.0$\times10^{-6}$ & $\mathrm{kgCOD/m^{3}}$ \\
$Y_{\mathrm{H_{2}}}$        & 0.06                        & $\mathrm{kgCOD_X/kgCOD_S}$                       \\
$k_{\mathrm{dec,i}}$        & 0.02                        & $\mathrm{d^{-1}}$      \\
$K_{I,\mathrm{H_{2}}}$      & 3.5$\times10^{-6}$ & $\mathrm{kgCOD/m^{3}}$ \\ \hline
\end{tabular}
\caption{Parameters used in the two-tiered food chain and three-tiered food web models, where d represents days, and $\mathrm{COD_X}$ and $\mathrm{COD_S}$ are the Chemical Oxygen Demand (COD) of the biomass and substrate, respectively.  The derivation of these parameter values is given in~\ref{appendix:ParamEst}.\label{params}}
\end{table}

\subsection{Dimensionless form}
It is beneficial to scale unit-dependent equations to a dimensionless form; this significantly reduces the number of parameters describing the dynamics, thereby simplifying the subsequent analyses. Using the notation of Xu \textit{et al.}~\cite{xu11}, after Baltzis and Fredrickson~\cite{baltzis84}, the following dimensionless terms are defined

\begin{align}
& \tau = k_{m,\mathrm{ch}}Y_{\mathrm{ch}}t;  \nonumber \\
& s_{0} = \frac{S_{\mathrm{ch}}}{K_{S,\mathrm{ch}}}; \ s_{1} = \frac{S_{\mathrm{ph}}}{K_{S,\mathrm{ph}}}; \ s_{2} = \frac{S_{\mathrm{H_2}}}{K_{S,\mathrm{H_2}}}; \nonumber \\
& x_{0} = \frac{X_{\mathrm{ch}}}{K_{S,\mathrm{ch}}Y_{\mathrm{ch}}}; \ x_{1} = \frac{X_{\mathrm{ph}}}{K_{S,\mathrm{ph}}Y_{\mathrm{ph}}}; \ x_{2} = \frac{X_{\mathrm{H_2}}}{K_{S,\mathrm{H_2}}Y_{\mathrm{H_2}}}. \nonumber
\end{align}

With these transformations, the dynamical system of ODEs given in Equations (\ref{eqm2}-\ref{eqm5}) reduces to

\begin{align}
\label{eqd2} \frac{\mathrm{d}x_{0}}{\mathrm{d}\tau} & = g_1(x_0,s_0,s_2) = -\alpha x_{0}+\mu_{0}x_{0}-k_{A}x_{0} \\
\label{eqd4} \frac{\mathrm{d}x_{1}}{\mathrm{d}\tau} & = g_2(x_1,s_1,s_2) =  -\alpha x_{1}+\mu_{1}x_{1}-k_{B}x_{1} \\
\label{eqd6} \frac{\mathrm{d}x_{2}}{\mathrm{d}\tau} & = g_3(x_2,s_2) =  -\alpha x_{2}+\mu_{2}x_{2}-k_{C}x_{2} \\
\label{eqd1} \frac{\mathrm{d}s_{0}}{\mathrm{d}\tau} & = g_4(x_0,s_0,s_2) = \alpha(u_{f}-s_{0}) - \mu_{0}x_{0} \\
\label{eqd3} \frac{\mathrm{d}s_{1}}{\mathrm{d}\tau} & = g_5(x_0,x_1,s_0,s_1,s_2) \nonumber \\
& =  \alpha (u_{g}-s_{1}) + \omega_{0}\mu_{0}x_{0} -\mu_{1}x_{1} \\
\label{eqd5} \frac{\mathrm{d}s_{2}}{\mathrm{d}\tau} & = g_6(x_0,x_1,x_2,s_0,s_1,s_2) \nonumber \\
& =  \alpha (u_{h}-s_{2})-\omega_{2}\mu_{0}x_{0}+\omega_{1}\mu_{1}x_{1}-\mu_{2}x_{2},  
\end{align} 
where, for simplicity, the following parameters are introduced

\begin{align}
& \alpha = \frac{D}{k_{m,\mathrm{ch}}Y_{\mathrm{ch}}}; \ u_{f} = \dfrac{S_{\mathrm{ch,in}}}{K_{S,\mathrm{ch}}}; \ u_{g} = \dfrac{S_{\mathrm{ph,in}}}{K_{S,\mathrm{ph}}}; \ u_{h} = \dfrac{S_{\mathrm{H_{2},in}}}{K_{S,\mathrm{H_{2}}}}; \nonumber \\
& \omega_{0} = \dfrac{K_{S,\mathrm{ch}}}{K_{S,\mathrm{ph}}}\dfrac{224}{208}(1-Y_{\mathrm{ch}}); \ \omega_{1} = \dfrac{K_{S,\mathrm{ph}}}{K_{S,\mathrm{H_2}}}\dfrac{32}{224}(1-Y_{\mathrm{ph}}); \nonumber \\
& \omega_{2} = \dfrac{16}{208}\dfrac{K_{S,\mathrm{ch}}}{K_{S,\mathrm{H_2}}}; \nonumber \\
& \phi_{1} = \dfrac{k_{m,\mathrm{ph}}Y_{\mathrm{ph}}}{k_{m,\mathrm{ch}}Y_{\mathrm{ch}}}; \ \phi_{2} = \dfrac{k_{m,\mathrm{H_2}}Y_{\mathrm{H_2}}}{k_{m,\mathrm{ch}}Y_{\mathrm{ch}}}; \nonumber \\
& K_{P} = \dfrac{K_{S,\mathrm{H_2},c}}{K_{S,\mathrm{H_2}}}; \ K_{I} = \dfrac{K_{S,\mathrm{H_2}}}{K_{I,\mathrm{H_2}}}; \nonumber \\
& k_{A} = \dfrac{k_{\rm dec,\mathrm{ch}}}{k_{m,\mathrm{ch}}Y_{\mathrm{ch}}}; \ k_{B} = \dfrac{k_{\mathrm{dec,ph}}}{k_{m,\mathrm{ch}}Y_{\mathrm{ch}}}; \ k_{C} = \dfrac{k_{\rm dec,\mathrm{H_2}}}{k_{m,\mathrm{ch}}Y_{\mathrm{ch}}}; \nonumber \\
& \mu_{0}(s_0,s_2) = \dfrac{s_{0}}{1+s_{0}}\dfrac{s_{2}}{K_{P}+s_{2}}; \ \mu_{1}(s_1,s_2) = \dfrac{\phi_{1}s_{1}}{1+s_{1}}\dfrac{1}{1+K_{I}s_{2}}; \nonumber \\ 
& \mu_{2}(s_2) = \dfrac{\phi_{2}s_{2}}{1+s_{2}}. \nonumber
\end{align}

\subsection{Steady-states}

The steady-states of this system of ODEs are obtained by setting $g_i=0$ (for $i=1,\dots,6$) and solving simultaneously.  This predicts eight possible steady-states:
\begin{description}
\item[SS1:] The trivial steady-state where all three populations are washed out ($x_0=x_1=x_2=0$).

\item[SS2:] Only the methanogen population is maintained ($x_0 = x_1 = 0, x_2 \neq 0$).

\item[SS3:] The phenol degraders and methanogens are washed out ($x_0 \neq 0, x_1 = x_2=0$). 

\item[SS4:] The hydrogenotrophic methanogens are washed out while the chlorophenol and phenol degraders are maintained ($x_0 \neq 0, x_1 \neq 0, x_2 = 0$). 

\item[SS5:] Only the phenol degraders are washed out ($x_0 \neq 0, x_1 = 0, x_2 \neq 0$). 

\item[SS6:] All three populations are present ($x_0 \neq 0, x_1 \neq 0, x_2 \neq 0$). 

\item[SS7:]  Only the phenol degraders are present ($x_0 = 0, x_1 \neq 0, x_2 = 0$). 

\item[SS8:]  The phenol degraders and methanogens are maintained in the system ($ x_0 = 0, x_1 \neq 0, x_2 \neq 0$). 
\end{description}

For full chlorophenol mineralisation and, as such, the only desired operating condition, SS6 must be stable.

\subsection{Stability analysis}

The challenge of analytically characterising dynamical systems of ODEs is well understood. Typically, for systems of high dimensions, it is intractable to finding explicit solutions and one must resort to obtaining solutions numerically. For example, the Routh-Hurwitz theorem allows for an explicit analysis of the stability of steady-states, but is intractable beyond five dimensions~\cite{may73} and is excluded here. Relying on numerical solutions, however, is problematic as the extent and resolution of the results are limited by the choice of model parameters and the computational resources. Nevertheless, here analytical expressions may be found for some of the steady-states that can inform general rules about their viability. To present the results of the stability analysis coherently, all steady-states are solved numerically for a realistic range of operational and kinetic parameters with a suitable resolution of $5000\times5000$ solution points. A brief discussion of the numerical methods used in this analysis is provided in~\ref{appendix:numerics}.

In order that a steady-state be meaningful, all variable concentrations must be non-negative, while Eq.~\eqref{eqm1} also gives the condition that $S_{\mathrm{ch}} < S_{\mathrm{ch,in}}$, (or $s_0 < u_f$ in dimensionless form). For model extensions with phenol and hydrogen addition, the conditions that $S_{\mathrm{ph}} < S_{\mathrm{ph,in}}$ ($s_1 < u_g$) and $S_{\mathrm{H_2}} < S_{\mathrm{H_2,in}}$ ($s_2 < u_h$) are also necessary. It is well established that the stability of a system of autonomous ordinary differential equations (ODEs) can be determined by investigating the eigenvalues of the corresponding Jacobian matrix~\cite{pimm02}. The Jacobian for the system of Equations (\ref{eqd2}-\ref{eqd5}) corresponds to the $6\times6$ matrix 

\begin{equation}
J = \left[\scalemath{1}{ \begin{array}{cccccc}
\dfrac{\partial g_1}{\partial x_0} & 0 & 0 & \dfrac{\partial g_1}{\partial s_0} & 0 & \dfrac{\partial g_1}{\partial s_2} \\
0 & \dfrac{\partial g_2}{\partial x_1} & 0 & 0 & \dfrac{\partial g_2}{\partial s_1} & \dfrac{\partial g_2}{\partial s_2} \\
0 & 0 & \dfrac{\partial g_3}{\partial x_2} & 0 & 0 & \dfrac{\partial g_3}{\partial s_2} \\
\dfrac{\partial g_4}{\partial x_0} & 0 & 0 & \dfrac{\partial g_4}{\partial s_0} & 0 & \dfrac{\partial g_4}{\partial s_2} \\
\dfrac{\partial g_5}{\partial x_0} & \dfrac{\partial g_5}{\partial x_1} & 0 & \dfrac{\partial g_5}{\partial s_0} & \dfrac{\partial g_5}{\partial s_1} & \dfrac{\partial g_5}{\partial s_2} \\
\dfrac{\partial g_6}{\partial x_0} & \dfrac{\partial g_6}{\partial x_1} & \dfrac{\partial g_6}{\partial x_2} & \dfrac{\partial g_6}{\partial s_0} & \dfrac{\partial g_6}{\partial s_1} & \dfrac{\partial g_6}{\partial s_2} \\ \end{array} }\right].
\end{equation}

Note that for the two-tier model, for which $X_{\rm ch}=S_{\rm ch}=0$, the Jacobian reduces to a $4\times4$ matrix.  The Jacobian is then evaluated at a given steady-state, denoted $J_{\rm SSi}$ (i=1,...,8), and its eigenvalues calculated.  If the real parts of all the eigenvalues obtained from the Jacobian are negative, then the state is stable. If one or more of the eigenvalues have a positive real part then the steady-state is unstable.

In Section~\ref{Results}, a thorough investigation is carried out in order to determine when each steady-state is viable. Firstly, a two-tier model describing only phenol degradation is considered, where $x_0=s_0=0$, reducing the system of ODEs to four equations. Subsequently, the full model for chlorophenol mineralisation is studied. 

\section{Results}\label{Results}

\subsection{Two-tier phenol model}

\begin{figure}
  \center
    \includegraphics[width=0.48\textwidth,clip]{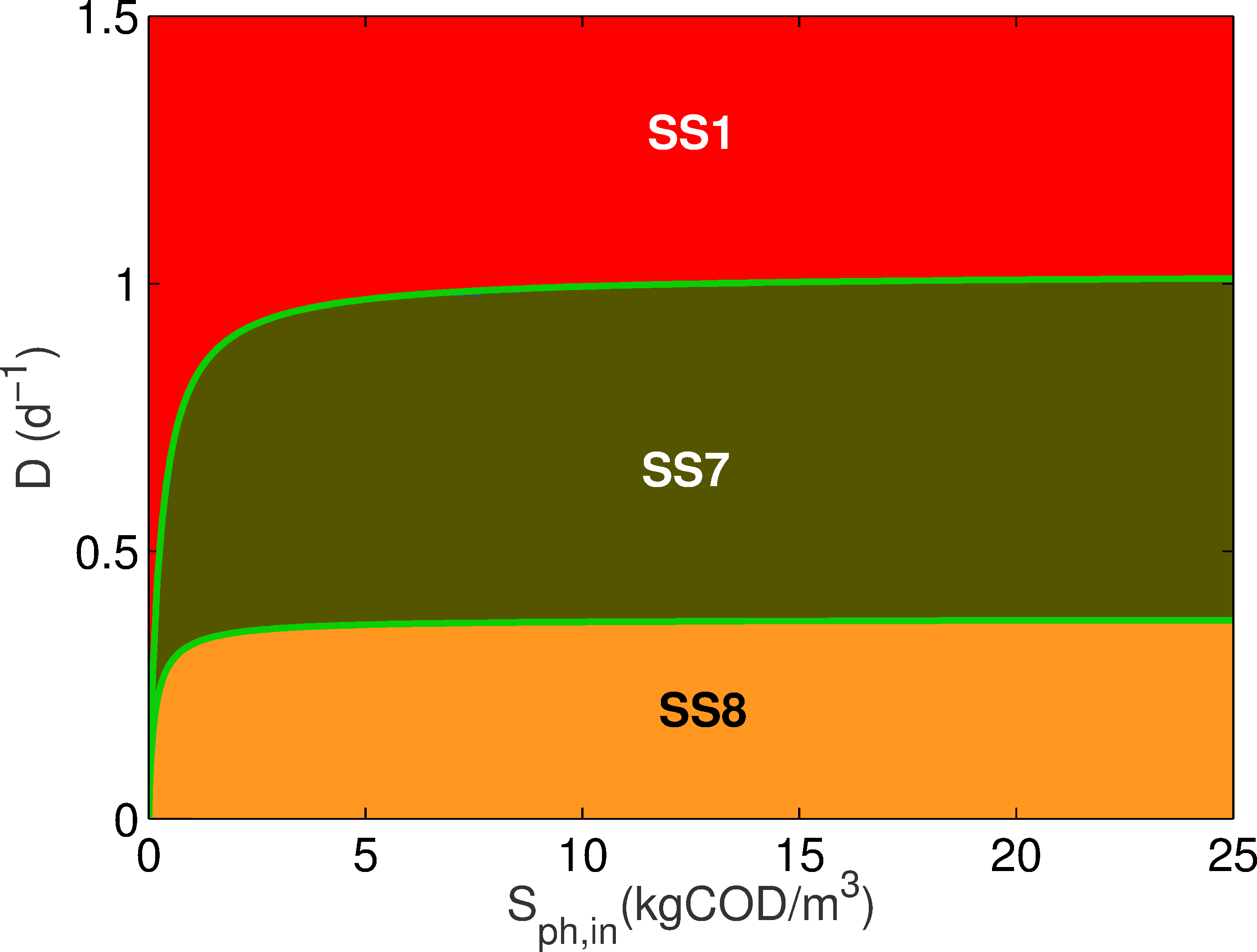}
      \caption{Steady-state diagram for operational parameters $D$ and $S_{\mathrm{ph,in}}$, in the two-tier phenol model. Green lines between steady-states: explicit functions of the  steady--state partitions, after Xu \textit{et al.}~\cite{xu11}.\label{phenola}}
\end{figure}

An initial analysis was performed on the two-species model feeding on phenol. As expected from the model structure, the results are similar to those shown for a propionate-degrading bi-culture with a maintenance term ($k_{\rm dec}$) ~\cite{xu11}, where three stable steady-states emerge: SS1, SS7 and SS8. Figure~\ref{phenola} shows the steady-state diagram for the phenol model, demonstrating that the system has three mutually exclusive stable steady-states under the range of parameters and operating conditions chosen. Numerical simulations under conditions producing SS7 (not shown) indicated that the concentration of phenol degraders are comparatively low (compared to SS8 populations), resulting in low hydrogen production. This in turn results in the washout of the methanogen, whilst the phenol degraders, under significant phenol concentrations, can be maintained up to dilution rates equivalent to their theoretical maximum growth rate minus the decay constant, $\mu_{{\rm max},\mathrm{ph}} - k_{\rm dec,\mathrm{ph}} = 1.02 \mathrm{d^{-1}}$. Although numerical analysis of the steady-states was performed here, it is possible to get explicit quadratic functions of the parameter pair ($S_{\mathrm{ph,in}}$, $D$) for the steady-state partitions. Using the method described in Xu \textit{et al.}~\cite{xu11}, the functions for the partitions are (in non-dimensionless form) 

\begin{align}
F_1(S_{\mathrm{ph,in}},D) =& S_{\mathrm{H_{2},\beta}}^2 - \left[\gamma\left(S_{\mathrm{ph,in}}+K_{S,\mathrm{ph}}\right) \right. \nonumber \\ & \left. 
+\frac{K_{I,\mathrm{H_2}}}{K_{S,\mathrm{H_2}}} \left(\frac{k_{m,\mathrm{ph}}Y_{\mathrm{ph}}}{D+k_{\rm dec,\mathrm{ph}}}-1\right)\right]S_{\mathrm{H_{2},\beta}} \nonumber \\  + \gamma \frac{K_{I,\mathrm{H_2}}}{K_{S,\mathrm{H_2}}}&\left(\frac{S_{\mathrm{ph,in}}k_{m,\mathrm{ph}}Y_{\mathrm{ph}}}{D+k_{\rm dec,\mathrm{ph}}}-S_{\mathrm{ph,in}}-K_{S,\mathrm{ph}}\right) \\
F_2(S_{\mathrm{ph,in}},D) =& \frac{S_{\mathrm{ph,in}}}{K_{S,\mathrm{ph}}+S_{\mathrm{ph,in}}}-\left(\frac{D + k_{\rm dec,\mathrm{ph}}}{k_{m,\mathrm{ph}}Y_{\mathrm{ph}}}\right),
\end{align}
where
\begin{align}
S_{\mathrm{H_{2},\beta}} &= \frac{1}{\dfrac{k_{m,\mathrm{H_{2}}}Y_{\mathrm{H_{2}}}}{D+k_{\rm dec,\mathrm{ph}}}-1}, \ \gamma = \frac{32}{224}\frac{\left(1-Y_{\mathrm{ph}}\right)}{K_{S,\mathrm{H_2}}}. \nonumber
\end{align}

where $ F_1=0$ relates to methanogen washout and $F_2=0$ to phenol degrader washout. These partitions are shown by the green lines in Fig.~\ref{phenola}.

Having observed that the phenol model is always stable and the desired operational state (SS8) is bounded asymptotically at higher phenol input concentrations, the next sections explore the properties of the model when extended to a three species dechlorinating tri-culture with hydrogen cycling.

\subsection{Three-tier chlorophenol model}

Solutions were found for the three steady-states identified (SS1, SS4 and SS6). Defining

\begin{align}
&E = \frac{\partial \mu_{0}}{\partial s_{0}} > 0; \ F = \frac{\partial \mu_{0}}{\partial s_{2}} > 0; \ G = \frac{\partial \mu_{1}}{\partial s_{1}} > 0; \nonumber \\
&H = \dfrac{\partial \mu_{1}}{\partial s_{2}} < 0;  \ I = \dfrac{\partial \mu_{2}}{\partial s_{2}} > 0,
\end{align}
then the Jacobians for SS1, SS4 and SS6, respectively, are

\[J_{\rm SS1} = \left[ {\begin{array}{@{}cccccc@{}}
-\alpha+\mu_{0}-k_{A} & 0 & 0 & 0 & 0 & 0 \\
0 & -\alpha +\mu_1- k_{B} & 0 & 0 & 0 & 0 \\
0 & 0 & -\alpha+\mu_{2}-k_{C} & 0 & 0 & 0 \\
-\mu_{0} & 0 & 0 & -\alpha & 0 & 0 \\
\omega_{0}\mu_{0} & -\mu_1 & 0 & 0 & -\alpha & 0 \\
-\omega_{2}\mu_{0} & \omega_1\mu_1 & -\mu_{2} & 0 & 0 & -\alpha \end{array}} \right]\] 

\[J_{\rm SS4} = \left[\scalemath{0.635}{\begin{array}{@{}cccccc@{}}
0 & 0 & 0 & Ex_{0} & 0 & Fx_{0} \\
0 & 0 & 0 & 0 & Gx_{1} & Hx_{1} \\
0 & 0 & -\alpha+\mu_{2}-k_{C} & 0 & 0 & 0 \\
-(\alpha+k_{A}) & 0 & 0 & -\alpha-Ex_{0} & 0 & -Fx_{0} \\
\omega_{0}(\alpha+k_{A}) & -(\alpha+k_{B})  & 0 & \omega_{0}Ex_{0} & -\alpha-Gx_{1} & \omega_{0}Fx_{0}-Hx_{1} \\
-\omega_{2}(\alpha+k_{A}) & \omega_{1}(\alpha+k_{B})  & -\mu_{2} & -\omega_{2}Ex_{0} & \omega_{1}Gx_{1} & -\alpha-\omega_{2}Fx_{0}+\omega_{1}Hx_{1} \end{array}} \right]\] 

\[J_{\rm SS6} = \left[\scalemath{0.61}{\begin{array}{@{}cccccc@{}}
0 & 0 & 0 & Ex_{0} & 0 & Fx_{0} \\
0 & 0 & 0 & 0 & Gx_{1} & Hx_{1} \\
0 & 0 & 0 & 0 & 0 & Ix_{2} \\
-(\alpha+k_{A}) & 0 & 0 & -\alpha-Ex_{0} & 0 & -Fx_{0} \\
\omega_{0}(\alpha+k_{A}) & -(\alpha+k_{B})  & 0 & \omega_{0}Ex_{0} & -\alpha-Gx_{1} & \omega_{0}Fx_{0}-Hx_{1} \\
-\omega_{2}(\alpha+k_{A}) & \omega_{1}(\alpha+k_{B})  & -(\alpha+k_{C}) & -\omega_{2}Ex_{0} & \omega_{1}Gx_{1} & -\alpha-\omega_{2}Fx_{0}+\omega_{1}Hx_{1}-Ix_{2} \end{array}} \right].\]

A full analysis of the stability for each steady-state is provided in \ref{appendix: full_model}. By inspecting the eigenvalues ($\lambda_{i}, i=1,\dots,6$) of the Jacobian matrix for each steady-state, the following conditions for stability are found:

\begin{description}
\item[SS1:] Meaningful and stable, with $s_0=u_f$.

\item[SS4:] Meaningful when the roots of the cubic function (Eq.~\eqref{eq:b11}) for $s_0$ are positive and the conditions $\alpha < \left(s_0/(1+s_0)\right) - k_A$ and $\alpha < \left(\phi_1/(1+K_Is_2)\right) - k_B$ are met. Additionally, it can be seen from Eq.~\eqref{eq:b18} that $x_0$ can only be meaningful if $\omega_0\omega_1 > \omega_2$. Solving this inequality results in the condition $(1-Y_{\mathrm{ch}})(1-Y_{\mathrm{ph}}) > 0.5$, such that $x_0$ is meaningful provided $0<Y_{\mathrm{ph}}<0.5$ and $0<Y_{\mathrm{ch}}<0.5$. Stability requires that $\alpha > \mu_{2} - k_{C}$ and $\mathbb{R}(\lambda_{i})<0$ ($i=2,\dots,6$).

\item[SS6:] Meaningful for the following conditions

\begin{align}
\nonumber& u_{f} > s_0, ~\alpha < \phi_2 - k_C, \ \alpha < \frac{\phi_{1}}{1+K_{I}s_{2}}-k_B, \\ \nonumber&\alpha < \frac{\omega_{0}k_{A}x_{0}}{s_{1}-\omega_{0}x_{0}}, 
 \ \alpha < \frac{\phi_1}{K_P+s_2}-k_A, \\ \nonumber& \alpha < \frac{\omega_{2}k_{A}x_{2}-\omega_{1}k_{B}x_{1}}{\omega_{1}x_{1}-\omega_{2}x_{2}-s_{2}}, 
\end{align}
and stable when $\mathbb{R}(\lambda_{i})<0$ ($i=1,\dots,6$).
\end{description}

\begin{figure}
    \includegraphics[width=0.47\textwidth,clip]{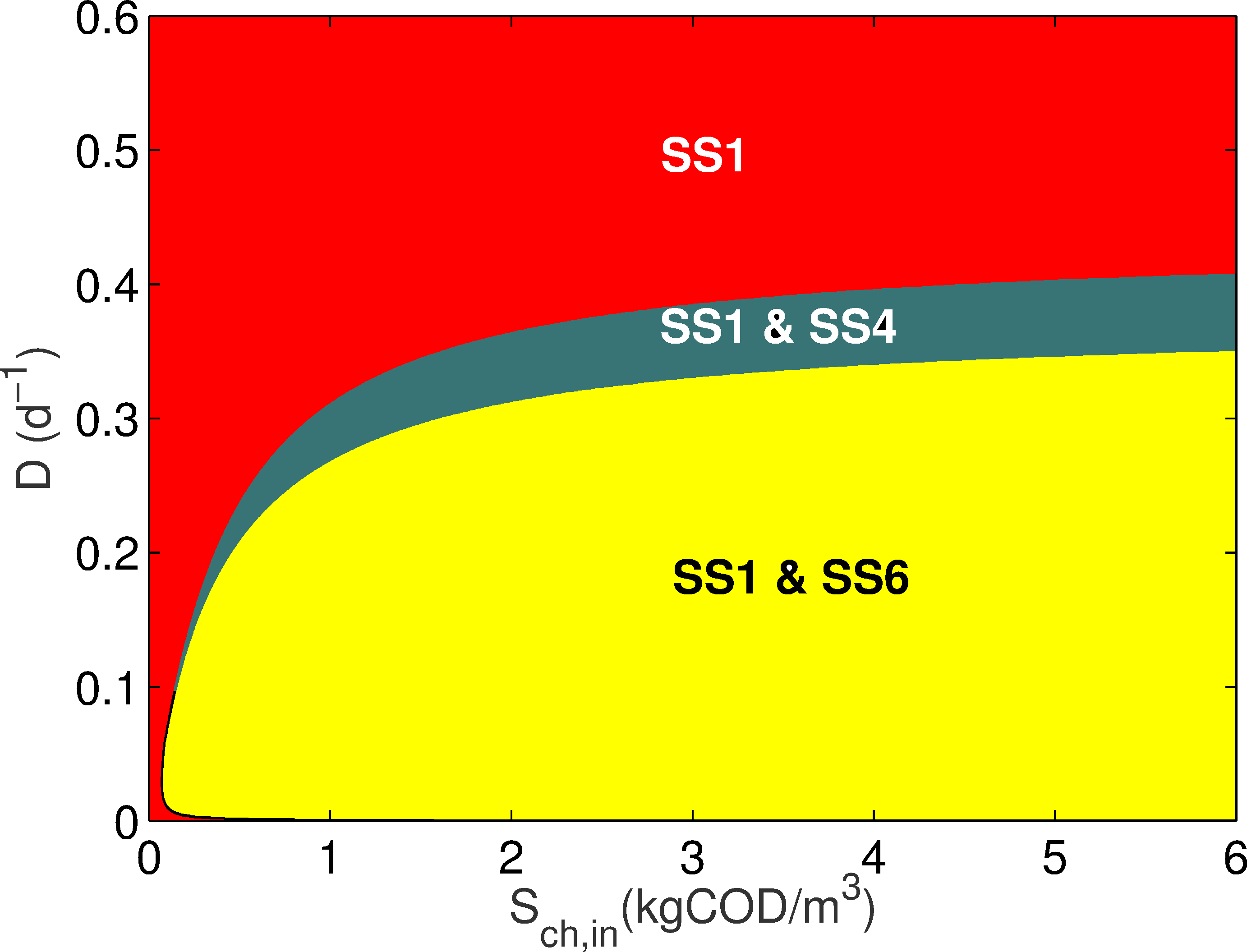}
      \caption{Steady--state diagram for operational parameters $D$ and $S_{\mathrm{ch,in}}$ in the three-tier chlorophenol model ($S_{\mathrm{ph,in}}=S_{\mathrm{H_2,in}}=0$).}
  \label{sm_num}
\end{figure}

Numerical simulation of the model was performed using the parameters listed in Table~\ref{params}, over a range of operating conditions ($\alpha$,$u_f$). The stable steady-state regions are shown in Fig.~\ref{sm_num}. As determined from the stability analysis, SS1 is meaningful under all operating conditions and bistable steady-states with SS4 and SS6 are found in specific regions within the numerical limits of the simulation, depending on the initial conditions. An interesting phenomenon not observed in previous two-tiered models is the existence of SS1 at low dilution rates but increasing chlorophenol input. It is likely that the condition $u_f > s_0$ is not satisfied in this region and suggests that the system does not produce enough metabolites to sustain the syntrophic populations at such low substrate concentrations and flow rates.

\subsection{Three-tier chlorophenol model with hydrogen addition}

Given that the model under analysis includes the possibility of adding extraneous substrates (phenol and hydrogen) to the system, supplementing the flux produced by the biomass, it is fairly straightforward to extend its mathematical analysis with these terms included. It is hypothesised that the addition of hydrogen to the system will result in the retention of the methanogen at dilution rates up to its maximum growth rate ($\mu_{\rm max}$). It is anticipated that this leads to the extension of full chlorophenol mineralisation by allowing the establishment of a methanogenic population that can be maintained given the availability of hydrogen in high enough concentrations. 

When hydrogen addition is included, the hydrogen input term, $S_{\mathrm{H_{2},in}} > 0$ in Eq.~\eqref{eqm5}, and the dimensionless form, $u_{h} > 0$ in Eq.~\eqref{eqd5}. A further three steady-states are defined with this model; SS2, SS3 and SS5.

Following the same approach as for the three-tier chlorophenol model, six Jacobian matrices were derived for each steady-state, and their eigenvalues $\lambda_{i}$ (where $i=1,\dots,6$) found. In this case, the conditions for stability for each steady-state are as follows

\begin{description}
\item[SS1:] Meaningful and stable providing 

\begin{align}
& \frac{u_{f}}{1+u_{f}}\frac{u_{h}}{K_{P}+u_{h}}-k_{A} < \alpha; \ \frac{\phi_{2}u_{h}}{1+u_{h}} - k_{C} < \alpha. \nonumber
\end{align}
Given that, in this model, $k_{A} = k_{C}$, then for $\phi_{2} > 1$, the second condition becomes the prevailing factor in determining stability, and can be reduced to $\mu_{\rm max,\mathrm{H_{2}}} - k_{C} < \alpha$, where $\mu_{\rm max,\mathrm{H_{2}}} = k_{m,\mathrm{H_{2}}}Y_{\mathrm{H_{2}}}/k_{m,\mathrm{ch}}Y_{\mathrm{ch}}$.

\item[SS2:] Meaningful when $\alpha < \phi_2 - k_C$ and $u_h > s_2$ and stable when $K_{P}(u_{f}+1)(\phi_{2} - k_{C}) + k_{C} - u_{f} + k_{C}u_{f}/(K_{P}-1)(u_{f}+1)<\alpha$. Here it can be noted that the stability of SS2 does not depend on $u_{h}$, and as such will remain fixed for any hydrogen input concentration.

\item[SS3:] Meaningful given the roots of the quadratic function (Eq.~\eqref{eq:b3}) for $s_0$ are positive, $u_f > s_0$ and $u_h > \omega_2(u_f-s_0)$. For steady-state stability, it is necessary that $\alpha > \mu_1 - k_B$, $\alpha > \mu_{2} - k_{C}$ and for $\mathbb{R}(\lambda_{i})<0$ ($i=4,5,6$). Here, the stability of SS3 is dependent on both $u_{f}$ and $u_{h}$. The first condition dictates that at higher $u_f$, a greater amount of $u_h$ is also required to maintain stability, whereas the second condition indicates that at a fixed $u_f$, increasing $u_h$ will increase the dilution rate under which SS3 is stable up to a boundary described by a function $f(u_f,\mu_{\rm max,\mathrm{H_2}} - k_C)$.

\item[SS4:] Meaningful given the roots of the cubic function (Eq.~\eqref{eq:b11}) for $s_0$ are positive and the conditions $\alpha < \left(s_0/(1+s_0)\right) - k_A$, $\alpha < \left(\phi_1/(1+K_Is_2)\right) - k_B$, $u_h > \omega_1s_1 + s_2$ and $x_1 > 0$ are met. Stability is assured when $\alpha > \mu_2-k_C$ and for $\mathbb{R}(\lambda_{i})<0$ ($i=2,\dots,6$). As $u_h$ affects $s_0$ through the highly non-linear cubic expression, it is difficult to draw any conclusion about its effect on stability by analytical means.

\item[SS5:] Meaningful when the conditions $\phi_2 > \alpha + k_C$, $s_2 > (K_P+s_2)(\alpha+k_A)$, $u_f > s_0$ and $u_h > s_2+\omega_2 + (\omega_2k_Ax_0/\alpha)$ are met. Stability of this steady-state is assured when $\alpha > \mu_1-k_B$ and $\mathbb{R}(\lambda_{i})<0$ ($i=3,\dots,6$).

\item[SS6:] Meaningful when $\alpha < \phi_2 - k_C$, $\alpha < \left(\phi_1/(1+K_Is_2)\right)-k_B$, $\alpha < \left(\phi_1/(K_P+s_2)\right)-k_A$, $u_f>s_0$, $\alpha < \omega_0k_Ax_0/(s_1-\omega_0x_0)$ and $x_2 > 0$. SS6 is stable when $\mathbb{R}(\lambda_{i})<0$ ($i=1,\dots,6$). The analysis shows that SS6 stability is affected by hydrogen addition.

\end{description}

The results from the stability analysis are visualised in Fig.~\ref{whydr} for different concentrations of hydrogen addition. As can be seen, four steady-states (SS3-6) are affected by increasing concentrations of hydrogen. At very low hydrogen concentration addition (Fig.~\ref{whydr} (a)), SS3 is stable in the region between (0,0) and SS2 (SS1 is not shown as it remains fixed at $D = M_{\rm max,\mathrm{H_{2}}}-k_{\rm dec} = 2.08 \mathrm{d^{-1}}$, where $M_{\rm max}$ is the non-dimensionless form of $\mu_{\rm max}$). Moreover, at lower dilution rates, the system has two observable bistable conditions: 1) between SS3 and SS6 at the lowest dilution rates; and 2) between SS3 and SS4, before SS3 becomes the single stable steady-state at higher dilution rates. This can be rationalised as follows:

\begin{figure*}
\center
\subfigure{\includegraphics[width=0.49\textwidth,clip]{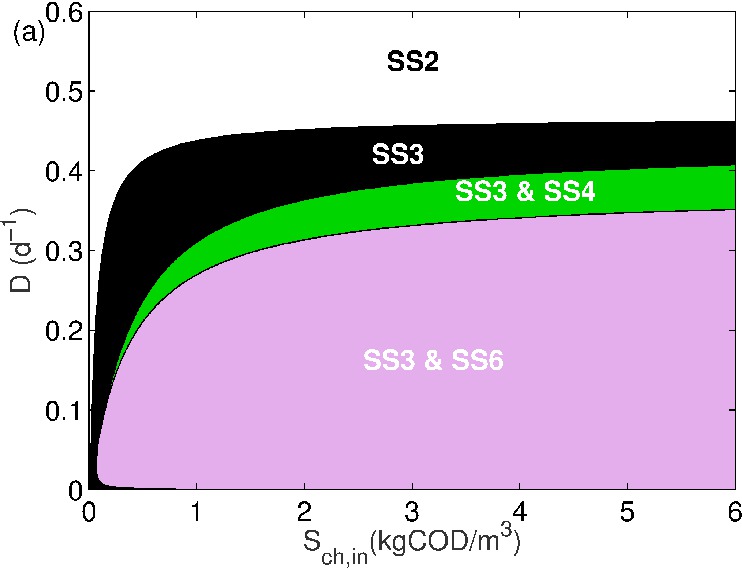}} 
\subfigure{\includegraphics[width=0.49\textwidth,clip]{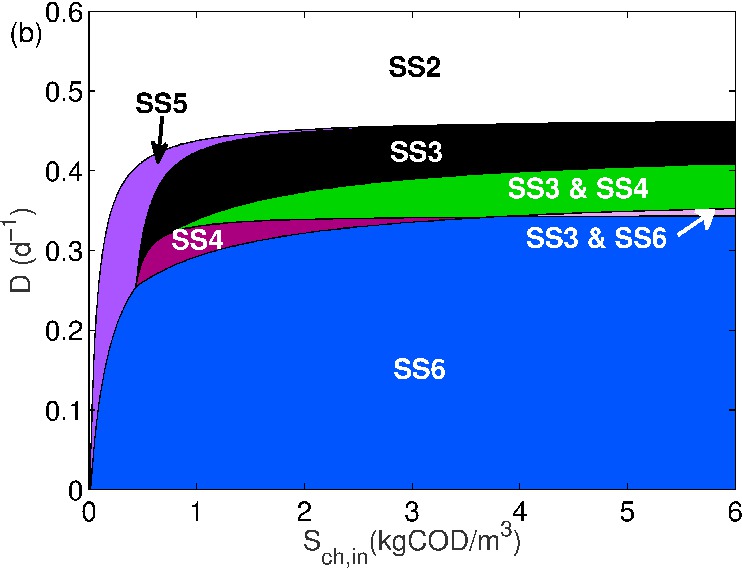}} 
\subfigure{\includegraphics[width=0.49\textwidth,clip]{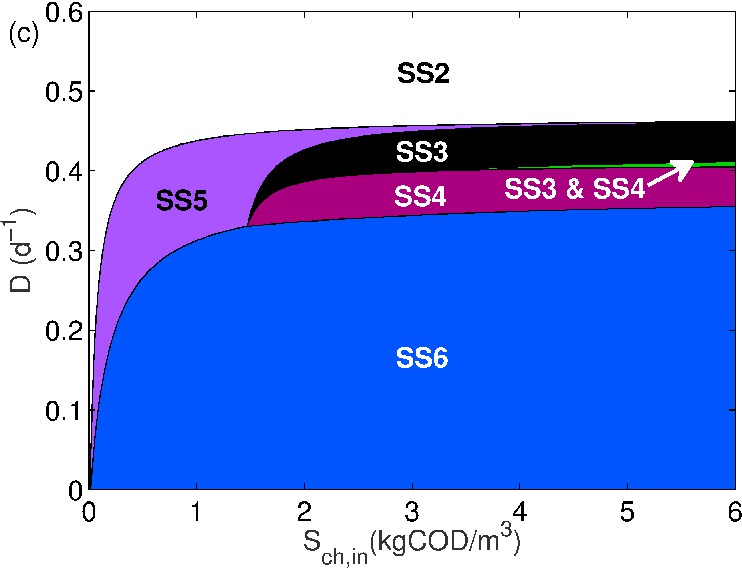}} 
\subfigure{\includegraphics[width=0.49\textwidth,clip]{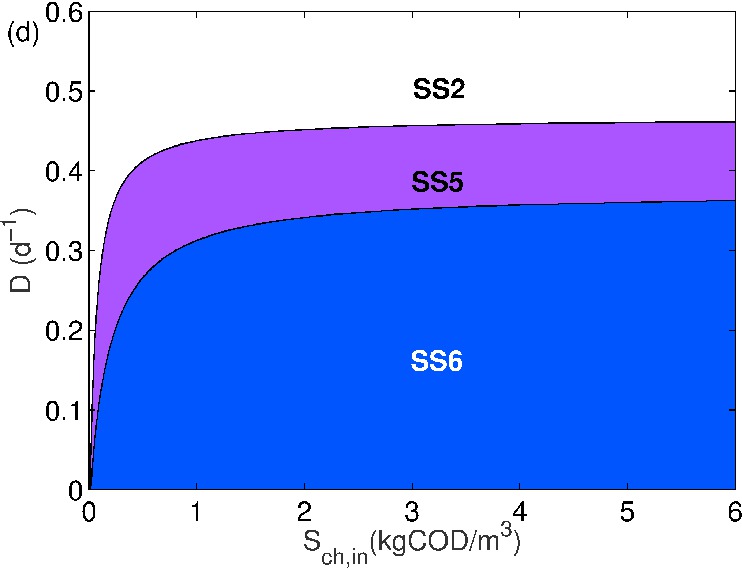}} 
\caption{Steady--state diagram for operational parameters $D$ and $S_{\mathrm{ch,in}}$ in the three-tier chlorophenol model with $\rm H_2$ addition, (a) $S_{\mathrm{H_2,in}}=2.67\times 10^{-5}$; (b) $S_{\mathrm{H_2,in}}=2.67\times 10^{-2}$; (c) $S_{\mathrm{H_2,in}}=10^{-1}$ and (d) $S_{\mathrm{H_2,in}}=2.67$ ($S_{\mathrm{ph,in}}=0$), (all $\rm kgCOD/m^3$).}
\label{whydr}
\end{figure*}

\begin{itemize}
\item The chlorophenol degrader can outcompete the methanogen for the additional hydrogen in the system. Depending on the initial biomass concentrations, this can result in complete washout of the phenol degrader and methanogen at low dilution rates. However, at higher chlorophenol concentrations, the system can be in stable equilibrium with all organisms present;
\item As the dilution rate increases, the methanogen is washed from the system as the chlorophenol degrader outcompetes it to become the sole hydrogen scavenger, as seen in Fig.~\ref{sm_num};
\item Eventually at higher dilution rates, the phenol degrader is washed out and full chlorophenol mineralisation is no longer possible. Nevertheless, beyond the theoretical $M_{\rm max,\mathrm{ch}}$, the methanogen is able to grow again (SS2), as there is a constant supply of substrate without an active competitor. 
\end{itemize} 

With increasing hydrogen addition interesting phenomena are observed (Fig.~\ref{whydr} (b)-(d)). Firstly, the bistability between SS3 and SS6 is replaced by a single steady-state, in which all organisms are present. Similarly, the SS3-SS4 bistability is replaced by only SS4, although this requires greater hydrogen concentrations at higher chlorophenol input. A new steady-state also begins to emerge (SS5), starting at low chlorophenol input, which eventually subsumes the other two varying steady-states (SS3 and SS4) at high hydrogen concentrations. The understanding from this is that at dilution rates higher than the upper boundary of SS6 and below the lower boundary of SS2, the phenol degrader is naturally washed out, but the presence of increasing amounts of hydrogen allows the methanogen to exist whilst in competition with the chlorophenol degrader. 

Most importantly, the extent of SS6 is also observed to increase below a specific $S_{\mathrm{ch,in}}$, with the additional hydrogen resource stabilising the methanogen population at higher dilution rates, limited now only by its maximum growth rate ($M_{\rm max,\mathrm{H_{2}}}$). Above this dilution rate, as with the three-tiered model, the production of phenol is limited by the stoichiometry of chlorophenol degradation. This restriction results in a limited supply of phenol under these operating conditions, which cannot produce enough growth in the phenol degrader to sustain its population. However, the increasing abundance of additional hydrogen leads to the establishment of a two-species trophic level with no intermediary organism (SS5). Nevertheless, the ability to extend the stable region of SS6 with addition of hydrogen is of note, particularly for practitioners. 

\begin{figure*}
\center
{\includegraphics[width=1\textwidth,clip]{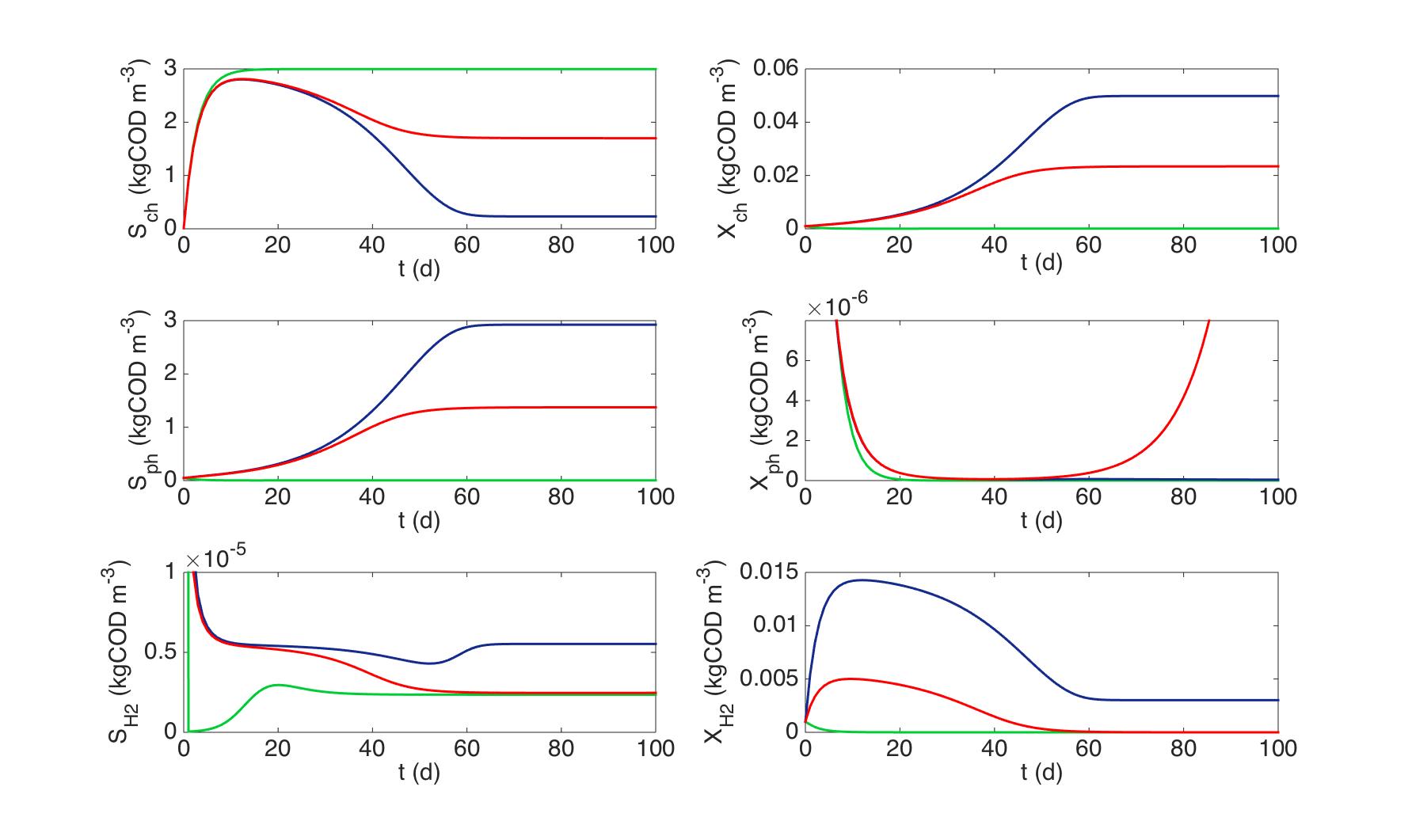}} 
\caption{Evolution of three-tier system with $H_2$ addition, where $D=0.36\ \mathrm{d^{-1}}$, $S_{\mathrm{ch,in}}=0.30$ and $S_{\mathrm{H_2,in}}=2.67\times 10^{-4}$ (green), $S_{\mathrm{H_2,in}}=1\times 10^{-1}$ (red), $S_{\mathrm{H_2,in}}=2.67$ (blue). Initial conditions: $X_{\mathrm{ch}}=10^{-3}$, $X_{\mathrm{ph}}=6.5\times 10^{-5}$, $X_{\mathrm{H_2}}=10^{-3}$, $S_{\mathrm{ch}}=10^{-2}$, $S_{\mathrm{ph}}=5\times 10^{-2}$, $S_{\mathrm{H_2}}=10^{-3}$ (all $\mathrm{kgCOD/m^3}$).}
\label{hydeff}
\end{figure*}

It has been shown that the inclusion of hydrogen addition in the three-tiered system can lead to an increase in the stable region for full chlorophenol mineralisation with regard to standard operating parameters. However, this is limited to a defined operating range. At lower $S_{\mathrm{ch,in}}$ there is washout of phenol degraders and establishment of methanogens at increasing hydrogen concentrations. It appears that a form of competitive exclusion occurs at certain ($D, S_{\mathrm{ch,in}}$). However, unlike the classical case, here the exclusion principle dictates that consumer A ($X_{\mathrm{ch}}$) is never excluded and either consumer B ($X_{\mathrm{H_{2}}}$) or producer A ($X_{\mathrm{ph}}$) are washed from the system. This can be seen in the system simulation shown in Fig.~\ref{hydeff}, which considers a single operational point ($D = \mathrm{0.36 d^{-1}}$, $S_{ch,in} = \mathrm{0.30 kgCOD/m^3}$), with increasing $S_{\mathrm{H_2,in}}$. 

At low hydrogen addition, the concentration is not enough to sustain the methanogen population whilst a small population of dechlorinators are present ($\mathrm{6\times10^{-5}kgCOD/m^3}$). The hydrogen not utilised by the chlorophenol degrader reaches an equilibrium that inhibits the phenol degraders and they are washed out of the system and, thus, SS3 is the dominant steady-state under these conditions for the initial conditions tested. As hydrogen addition is increased, an interesting property of the system emerges. Again, the chlorophenol degrader is able to utilise the extra hydrogen to produce more biomass and more phenol. However, the concentrations of additional hydrogen are not enough to sustain the methanogen population and the equilibrium concentration returns to a value approximate to the previous case. Whilst this exerts an inhibitory effect on the phenol degraders, the additional phenol substrate availability is enough to maintain a population in the system (steady-state not shown in Fig.~\ref{hydeff}), which switches the system to SS4. In the final case, hydrogen addition is further increased, allowing chlorophenol degraders to utilise most of the available chlorophenol and reach close to its maximum growth. As this growth is asymptotic, and because of the surplus hydrogen available, the methanogens are now able to utilise the remaining concentrations to maintain their population in the system. This results in an increased hydrogen equilibrium that exerts a greater inhibition on the phenol degraders, to such an extent that the additional phenol availability is not enough to avoid washout and the system moves to SS5.

It can be seen that this situation is fairly complex and is dependent on a number of factors. Principally, the equilibrium concentration of hydrogen is determined by the dilution rate, which in turn limits the growth of either producer A or consumer B.

\subsection{Three-tier chlorophenol model with phenol addition}

For such a non-linear system the addition of hydrogen could lead to deleterious and potentially unexpected behaviour, especially given its inhibitory effect on the phenol degraders.  It is therefore useful to examine the effect of adding phenol to the system, both as a sole exogenous substrate, but also in combination with hydrogen addition. In this way, it may be determined under what conditions, if any, full chlorophenol mineralisation can be extended in relation to the two operational parameters ($D$ and $S_{\mathrm{ch,in}}$). 

When phenol is added to the system the phenol inflow term, $S_{\mathrm{ph,in}}$ from Eq.~\eqref{eqm3} and $u_{g}$ from the dimensionless Eq.~\eqref{eqd3}, become positive. An additional two steady-states are defined with this model, SS7 and SS8.

The conditions for stability in each steady-state, determined in the same manner as previously described, are as follows:

\begin{description}
\item[SS1:] Stable for $\phi_{1}u_{g}/(1+u_{g}) - k_{B} < \alpha$. For high concentrations of phenol, this is equivalent to $\mu_{\rm max,\mathrm{ph}} - k_{B} < \alpha$, where  $\mu_{\rm max,\mathrm{ph}} = k_{m,\mathrm{ph}}Y_{\mathrm{ph}}/k_{m,\mathrm{ch}}Y_{\mathrm{ch}}$. However, at lower phenol concentrations, the effect of phenol lowers the minimum dilution rate under which SS1 is stable. 

\item[SS2:] Never stable as a contradiction occurs in the terms describing $s_2$ and $x_2$. For these variables to be meaningful, the conditions $\alpha - k_{C} < \phi_2$ and $\alpha - k_{C} > \phi_2$ must be satisfied, which is invalid (see Eqs.~\eqref{eq:b1} and~\eqref{eq:b2}).

\item[SS3:] Never stable as a contradiction occurs in the terms describing $s_2$ and $x_0$. With $u_h = 0$, $s_{0} > u_f$ for $s_2$ to be positive. This condition, however, results in $x_{0}$ always being negative and meaningless (see Eqs.~\eqref{eq:b4} and~\eqref{eq:b5}).

\item[SS4:] Meaningful given the roots of the cubic function (Eq.~\eqref{eq:b11}) for $s_0$ are positive and the conditions $\alpha < \left(s_0/(1+s_0)\right) - k_A$, $\alpha < \left(\phi_1/(1+K_Is_2)\right) - k_B$, $x_0 > 0$ and $x_1 > 0$ are met. Stability is assured when $\alpha > \mu_2-k_C$ and the roots of the characteristic polynomial (Eq.~\eqref{eq:SS5_poly}) have negative real parts. In the special case that hydrogen is also added to the system, it should be noted that if $u_g < s_1$, then the condition $u_h > s_2$ must be strictly observed. Similarly, if $u_h < s_2$, then $u_g$ must be greater than $s_1$.

\item[SS5:] Never stable as without $u_h$ then $x_2$ can never be positive (see Eq.~\eqref{eq:b29}).

\item[SS6:] Meaningful when $\alpha < \phi_2 - k_C$, $\alpha < \left(\phi_1/(1+K_Is_2)\right)-k_B$, $\alpha < \left(\phi_1/(K_P+s_2)\right)-k_A$, $u_f>s_0$, $x_0 > 0$ and $\alpha < (\omega_1x_1-\omega_2x_0)/s_2$. With these conditions met, stability of SS6 is guaranteed when the roots of the characteristic polynomial (Eq.~\eqref{eq:SS6_poly}) have negative real parts. In this case, it can be shown that SS6 is influenced by $u_g$.

\item[SS7:] Meaningful given the roots of the quadratic function for $s_1$ (Eq.~\eqref{eq:b49}) are positive and, subsequently, $\phi_{1}s_1/(1+s_1)K_I(\alpha+k_B) > 1/K_I$, and $u_g > s_1$. Therefore, stability is assured when $\alpha > \mu_0-k_{A}$, $\alpha > \mu_2-k_{C}$ and the roots of the characteristic polynomial (Eq.~\eqref{eq:poly_SS7}) have negative real parts. It can be seen that the stability of SS7 relies on $u_g$, but is also influenced by hydrogen if added to the system.

\item[SS8:] Stable when the following conditions are satisfied. Firstly, $s_2$ is meaningful given $\phi_2 > \alpha - k_{c}$. Stability is assured when $s_2$ is meaningful, $\alpha > \mu_0 - k_{A}$, and the roots of the characteristic polynomial (Eq.~\eqref{eq:poly_SS8}) have negative real parts. In general, for SS8 to be stable, $\alpha$ must be smaller than $\mu_{\mathrm{max, H_{2}}}$ and, therefore, $u_f$ should be small.
\end{description}

\begin{figure*}
\center
\subfigure{\includegraphics[width=0.49\textwidth,clip]{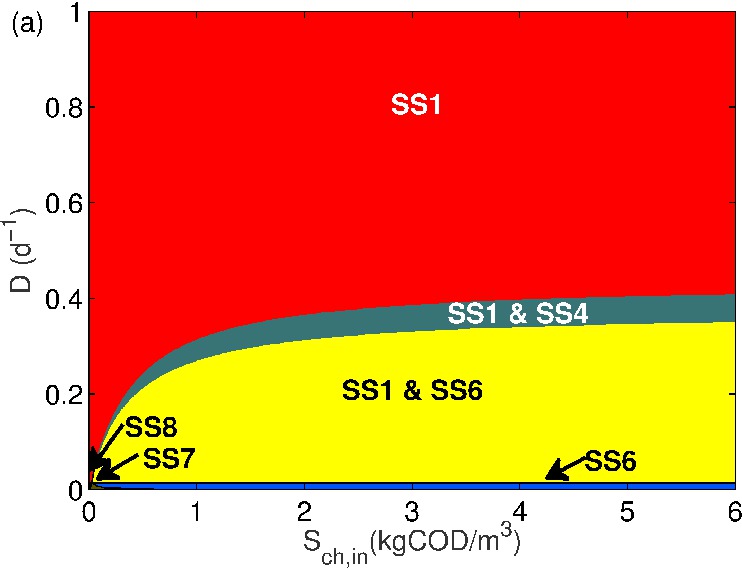}} 
\subfigure{\includegraphics[width=0.49\textwidth,clip]{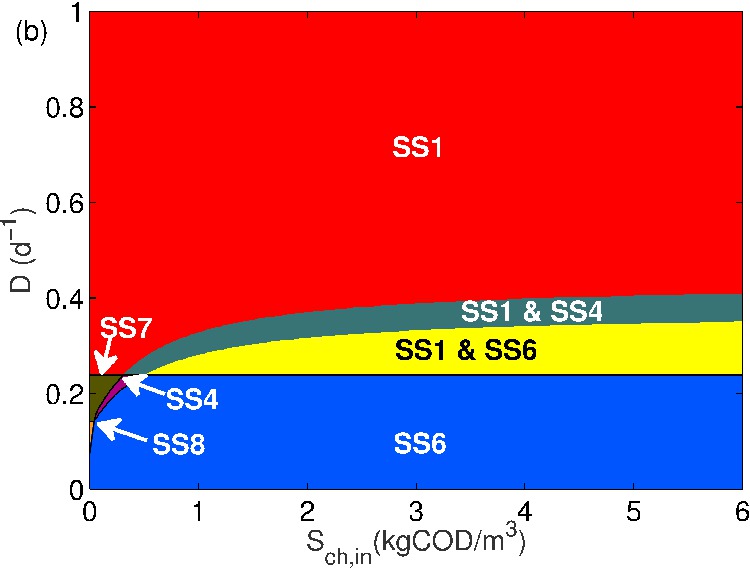}} 
\subfigure{\includegraphics[width=0.49\textwidth,clip]{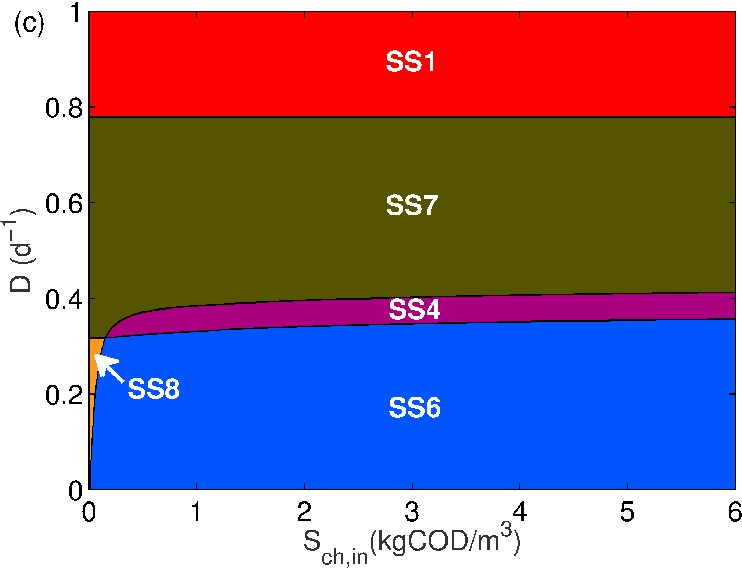}} 
\subfigure{\includegraphics[width=0.49\textwidth,clip]{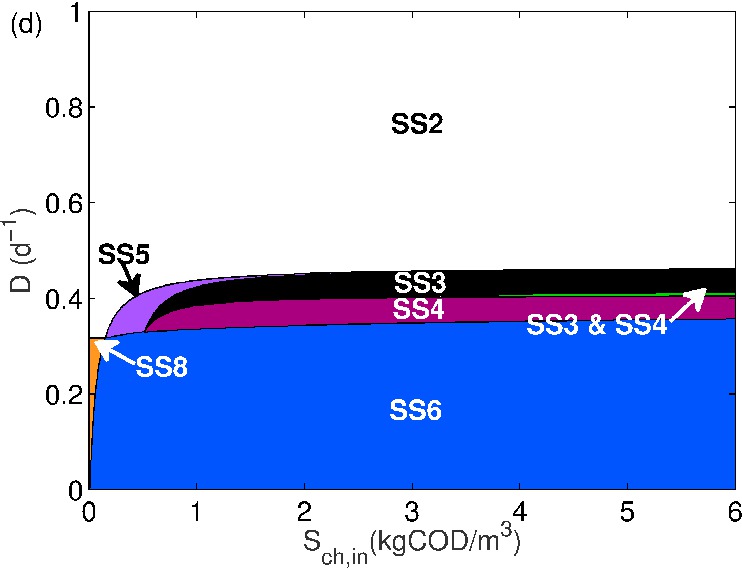}} 
\caption{Steady--state diagram for operational parameters $D$ and $S_{\mathrm{ch,in}}$ in the three-tier chlorophenol model with phenol addition, (a) $S_{\mathrm{ph,in}} = 1.00\times10^{-2}$ and $S_{\mathrm{H_{2},in}} = 0$; (b) $S_{\mathrm{ph,in}} = 1.00\times10^{-1}$ and $S_{\mathrm{H_{2},in}} = 0$; (c) $S_{\mathrm{ph,in}} = 1.00$ and $S_{\mathrm{H_{2},in}} = 0$; (d) $S_{\mathrm{ph,in}} = 1.00$ and $S_{\mathrm{H_{2},in}} = 2.67\times10^{-2}$ ($\mathrm{kgCOD/m^3}$).}
\label{wphen}
\end{figure*}

Again, the operational parameter plots are used to visualise the outputs from the stability analysis, as shown in Figs~\ref{wphen} (a)-(c). At low phenol addition, emergent steady-states (SS6-SS8) are observed at low dilution rates, with the bistability present in the standard three-tiered model comprising the rest of the system below the boundary for SS1. As phenol is increased, the two bistable regions disappear to be replaced by a single stable steady-state (SS4 and SS6). Both SS7 and SS8 also appear across a greater range of dilutions, although SS8 is confined to low chlorophenol addition, whilst SS8 eventually covers a region contained by the upper bounds of SS7 (low $S_{\mathrm{ch,in}}$) and SS4 (high $S_{\mathrm{ch,in}}$), and the boundary of SS1. Of greater interest is the observation that the extent of the desired condition, SS6, is greater for $S_{\mathrm{ch,in}}$ between approximately $0$ and $1\mathrm{kgCOD/m^3}$, replacing both SS4 and part of SS7.  

\subsection{Three-tier chlorophenol model with bi-substrate addition}

Figure~\ref{wphen} (d) shows an example when both hydrogen and phenol are added to the system. Here it can be seen that the two additional substrates each contribute to distinct emergent properties. The hydrogen addition allows for the maintenance of a methanogen population at higher dilution rates, whilst the phenol addition maintains the phenol degraders at lower dilution rates. Both additional substrates provide the conditions necessary for full chlorophenol mineralisation within a specific operational parameter space previously associated with methanogen washout. 

\begin{figure}
\center
\subfigure{\includegraphics[width=0.48\textwidth,clip]{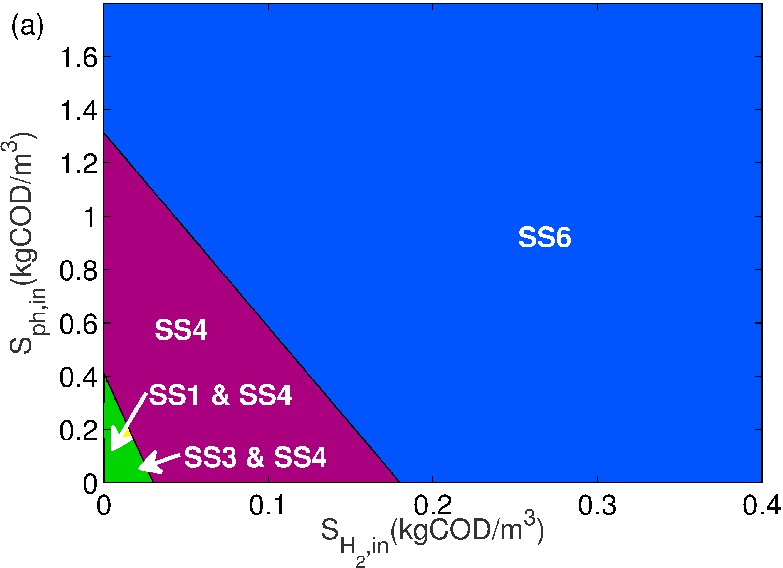}} 
\subfigure{\includegraphics[width=0.48\textwidth,clip]{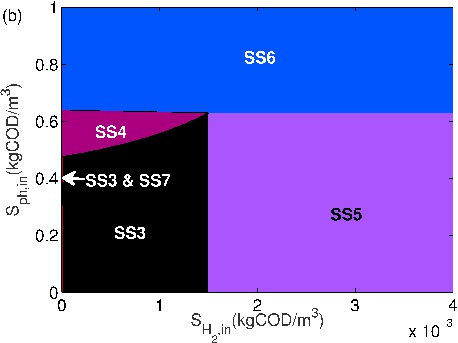}} 
\caption{Steady--state diagram for operational parameters $S_{\mathrm{ph,in}}$ and $S_{\mathrm{H_2,in}}$ in the three-tier chlorophenol model with (a) $D=\mathrm{0.25d^{-1}}$ and $S_{\mathrm{ch,in}}=\mathrm{0.5kgCOD/m^3}$; (b) $D=\mathrm{0.35d^{-1}}$ and $S_{\mathrm{ch,in}}=\mathrm{3kgCOD/m^3}$.}
\label{phvsh2}
\end{figure}

An extension of this analysis was undertaken by examining the effect of the two supplementary substrate additions on the maintenance of SS6 at higher dilution rates than observed in the standard three-tiered model. Figure~\ref{phvsh2} shows the steady-state plot for varying concentrations of hydrogen and phenol addition for two distinct operating conditions observed previously to give rise to SS6 with addition of an extraneous substrate. Figure~\ref{phvsh2} (a) demonstrates that both hydrogen and phenol addition can transform the system from undesirable bistability to a stable process with all three organisms present. Indeed, it is shown here that addition of either substrate can lead to SS6 under relatively high chlorophenol input conditions. 

An alternative situation under low chlorophenol addition is shown in Figure~\ref{phvsh2} (b). Here, it is clear that phenol addition at concentrations above $0.65\mathrm{kgCOD/m^3}$  provides the necessary conditions for full chlorophenol mineralisation, irregardless of hydrogen addition. Although it can also be seen that hydrogen addition does have some effect within a very small range of phenol concentration between about $0.63$ and $0.65\mathrm{kgCOD/m^3}$, allowing for stability of the methanogenic population and moving the steady-state from SS4 to SS6, it is relatively negligible given the influence of phenol with these operating parameters. 

\section{Bistability Analysis}

Throughout Section~\ref{Results}, numerous regions of bistability were observed between SS3-SS4 (Fig.~\ref{whydr} (a), (b) and (c)), SS3-SS6 (Fig.~\ref{whydr} (a) and (b)) and SS1-SS6 (Fig.~\ref{phenola}). In this Section, these bistable regions are investigated in order to provide a more in-depth description and understanding. Through analysis of these results, the steady-state having the larger basin of attraction is determined, i.e. the most likely outcome for a range of initial conditions. Note that the other four steady-states, even when meaningless or unstable, are still present mathematically and can impact on the overall dynamics of the system.

\begin{figure}
\center
  \includegraphics[width=0.48\textwidth,clip]{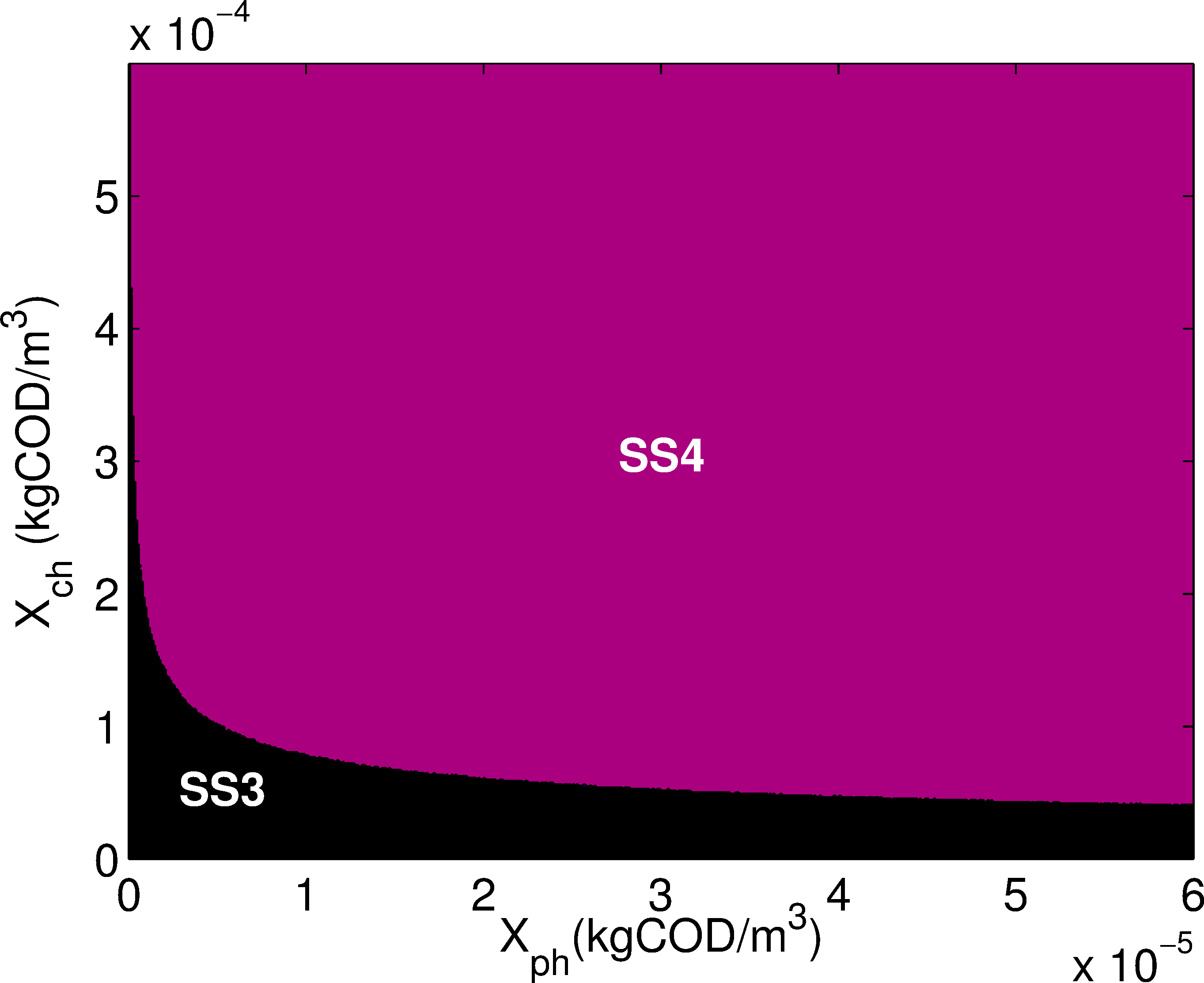}
  \caption{Phase plot showing final steady-state with initial conditions $X_{\mathrm{H_2}}=0$, $S_{\mathrm{ch}}=0.14$, $S_{\mathrm{ph}}=0.11$ and $S_{\mathrm{H_2}}=5.5\times 10^{-5}$ and parameters $D = \mathrm{0.375d^{-1}}, S_{\mathrm{ch,in}} = 3$ and $S_{\mathrm{H_2,in}} = 0.0267$ (All $X_i \ \mathrm{and}\ S_i \mathrm{\ in \ kgCOD/m^3)}$.}
  \label{bistability}
\end{figure}

Firstly, the region of bistability between SS3 and SS4 was considered. Figure~\ref{bistability} shows the final stable steady-states (SS3 or SS4) achieved for a range of initial conditions. Note that for SS3 or SS4, $X_{\mathrm{H_2}}=0$ and does not deviate from this value throughout, which simplifies the analysis required here. When $X_{\mathrm{ch}}$ is small, the system is likely to head towards SS3 where washout of $X_{\mathrm{ph}}$ occurs. Otherwise, the system achieves SS4. Very similar results were obtained for biologically meaningful substrate initial conditions between $1.3\times10^{-2}$ and $1.5\times10^{-1}$ for $S_{\mathrm{ch}}$, $9.0\times10^{-2}$ and $4.9\times10^{-1}$ for $S_{\mathrm{ph}}$ and $2.2\times10^{-12}$ and $4.7\times10^{-12}$ for $S_{\mathrm{H_2}}$ ($\mathrm{kgCOD/m^3}$). These results suggest that as long as sufficient dechlorinating biomass is present in the system, the most likely outcome is for both $X_{\mathrm{ch}}$ and $X_{\mathrm{ph}}$ to be present over time. Thus, the basin of attraction for SS4 is larger than that for SS3. In addition, changing the dilution rate $D$, $S_{\mathrm{ch,in}}$, or $S_{\mathrm{H_2,in}}$ within the bistable region once again leads to similar results.

A similar analysis of the bistable regions for SS3-SS6 and SS1-SS6 give a more dramatic outcome. If any of the three organisms, $X_\mathrm{ph}$, $X_\mathrm{ch}$ or $X_\mathrm{H_2}$, are initially zero, SS6 is not obtained. If all of these are present, even in tiny quantities, SS6 will be reached.
It is known that SS1 is always meaningful, thus a steady-state always occurs at the origin, be it stable or unstable. In the case of bistability, SS3-SS4 or SS3-SS6, the unstable steady-state is closer to SS3 rather than SS4 or SS6, respectively. This instability thus forces the dynamics away from the total washout and towards the state with the least washout.
This analysis highlights how not only the main parameters of the system can impact the final stable state, but also the importance of initial conditions. 

\section{Discussion}\label{Discussion}

In this work, a mathematical analysis of a three-tiered `food web' comprising three organisms with hydrogen addition and inhibition has been presented. Although mechanistic models of microbial interactions are somewhat ubiquitous, analytical approaches have been limited to simple two-species systems~\cite{ xu11, dubey00, wu04}. More recently, attempts at analysing more complete systems, such as the anaerobic digestion process, have been performed with some degree of rigour~\cite{bornhoft13, weedermann13}, however the practicality of these approaches is limited by the need for generalisation and increasing numbers of assumptions to allow for mathematical tractability. 

Although the three-tiered system is considered only a sub-process of anaerobic digestion and the analysis limited to a specific compound, the approach is general enough to provide information about the characteristics of such a process and allows for the possibility of extending the work to other systems and other compounds. Indeed the results shown here point to the fact that biological knowledge can inform mathematical approaches, whilst mathematical models can indicate or confirm biological properties of a system, succinctly and rationally. 

Here, four models were analysed by a combination of analytical and numerical techniques to obtain information regarding the extent and characteristics of system stability in a three-organism process anaerobically degrading chlorophenol. In this case, only the hydrogen part of the pathway was considered, with acetate being excluded from the analysis. The simple two-tiered food chain considering only phenol degradation with two species was shown to have the same characteristics observed as a previously studied system investigating propionate~\cite{xu11}; three steady-states and always stable. A second model introduced chlorophenol as the primary substrate and a dechlorinating organism forming a tri-culture that resulted in the emergence of bistable conditions between the complete washout steady-state and the two other viable states. Analysis of these bistabilities has shown that the desired operating condition (SS6) has the strongest basin of attraction such that complete washout can only occur when the initial condition for one of the biomass concentrations is zero. 

Although it is interesting to see the extent under which full chlorophenol mineralisation can occur using this standard model, of greater interest from an engineering perspective is the possibility of driving the system towards its limits of operational viability without compromising its function. With this in mind, the inclusion of additional input terms for hydrogen and phenol was undertaken, and corresponding stability analysis performed. In the case of hydrogen addition, it was expected that this would lead to a wider region of stability for the methanogen and, thus, extend the operational domain for SS6. Indeed, this was the case under relatively high concentrations of chlorophenol input, under which the methanogen population was maintained up to a theoretical maximum, dictated by its maximum growth rate. However at low chlorophenol input ($< 0.5\mathrm{kgCOD/m^3}$), the nature of the resource competition between the dechlorinator and methanogen is such that a switching of behaviour occurs in localities not conducive to stability of all three organisms. 

\begin{figure}
\center
  \includegraphics[width=0.48\textwidth,clip]{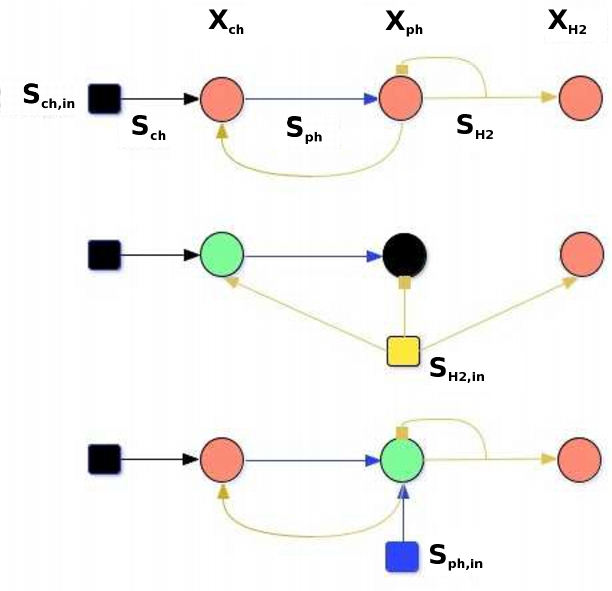}
  \caption{Schematic of observed microbial interactions in the three chlorophenol degradation models at low chlorophenol input concentrations. \textit{Top}: Three-tier chlorophenol model with hydrogen inhibition of the phenol degrader, \textit{Middle}: Hydrogen addition model with phenol degrader washout. \textit{Bottom}: Phenol addition model with extended stability. Organisms that are always present are shown as a green circle, a red circle is for when they are conditionally present, and a black circle represents biomass washout}
  \label{schematic}
\end{figure}

Under the standard three-tiered chlorophenol model, resource competition dictates that the dechlorinator is able to utilise the available hydrogen for growth more readily than the methanogen and under SS4 a competitive exclusion principle occurs and the latter is washed out. With hydrogen addition, however, an abundance of this resource allows the methanogen to access enough hydrogen to maintain a population at steady-state. However, below a threshold concentration and under specific dilution rates, an excess of hydrogen leads to inhibition of the phenol degrader and washout. In effect, the addition of hydrogen at this threshold results in a switching between SS4 and SS5, rather than a stabilisation at SS6. The addition of phenol, however, under specific operational conditions can result in the emergence of SS6 under low chlorophenol input conditions. Here, a specific form of competition between the dechlorinator and methanogen occurs (as with SS6 in the standard model), by which both organisms benefit from the production of hydrogen by the intermediate phenol degrader, whilst the phenol degrader is stabilised by the syntrophic hydrogen removal. In other words, the presence of the chlorophenol degrader allows for the production of phenol, which benefits both the phenol degrader and, indirectly, the methanogen. The presence of the methanogen reduces the inhibition on the phenol degrader, thus allowing the hydrogen resource to be maintained for both hydrogenotrophic populations. This can be seen as a form of mutualism rather than competition. The phenol addition plays a significant role under conditions where the chlorophenol degrader cannot produce enough phenol to maintain the phenol degraders, and under such conditions it is possible to achieve full chlorophenol mineralisation beyond the standard model system. A schematic of these models and their interactions for the low chlorophenol input condition is shown in Fig.~\ref{schematic}.

Although higher dimensional systems are analytically restricted and lack generality, the work reported here for a three-tiered 'food web' underlines the potential for applying localised stability analysis within meaningful operating and parameter ranges, to identify properties of the system that both increase fundamental understanding of the behaviour of such microbial systems, but also guide thought on ways to manipulate or control them for potential process improvement. Further work will include a thermodynamic rather than a kinetic inhibition term for the effect of hydrogen on the syntroph, and the extension of the model to polychlorinated phenols.

\section*{Acknowledgments}
This work was funded by the Biotechnology and Biological Sciences Research Council UK (BB/K003240/2 Engineering synthetic microbial communities for biomethane production) and by the Institute of Sustainability, Newcastle University.

\appendix

\section{Parameter estimation}\label{appendix:ParamEst}

The parameters of the model are based on the Chemical Oxygen Demand (COD) to be in consensus with the IWA Anaerobic Digestion Model No. 1~\cite{batstone02}. However, only those related to hydrogen can be found in the ADM1 (for high-rate mesophilic processes). Subsequently, the parameters for chlorophenol and phenol were chosen based on a combination of literature data and deduction based on their chemical oxidation reaction.

\paragraph{Parameters for chlorophenol}

Adrian \textit{et al.} used the \textit{Dehalococcoides mccartyi} strain CBDB1 to completely convert 2,3-dichlorophenol, all six trichlorophenols, all three tetrachlorophenols, and pentachlorophenol to lower chlorinated phenols~\cite{adrian07}, and their results are used partially here, with some deductive reasoning. The observed dechlorination rates in batch cultures with cell numbers of $10^{7}$mL$^{-1}$ was found to be 35 $\mu$Md$^{-1}$. The COD of the CBDB1 strain was first estimated using the empirical formula for bacterial cell $\mathrm{C_{5}H_{7}O_{2}N}$, with cell numbers of $10^{7}$mL$^{-1}$.

The oxidation reaction for $\mathrm{C_{5}H_{7}O_{2}N}$ is
\begin{align}
\mathrm{C_{5}H_{7}O_{2}N} + 5\mathrm{O_2} \rightarrow 5\mathrm{CO_2} + \mathrm{NH_3} + 2\mathrm{H_{2}O}.
\end{align}

From this equation, it is easy to show that 1 mole of $\mathrm{C_{5}H_{7}O_{2}N}$ requires 5 mole of $\mathrm{O_2}$, or $160$gCOD. The molar mass of $\mathrm{C_{5}H_{7}O_{2}N}$ is $113$g/mol. Therefore, the proportion of $\mathrm{C_{5}H_{7}O_{2}N}$ is $160/113 = 1.42$g. By applying this to general organic matter, a factor of $2$ is used. This means that each gram of biomass (e.g. strain CBDB1 here) requires $2$g of oxygen for growth. Subsequently, the cell number should be converted to  grams (dry weight) using a conversion factor of $1.2\times10^{-14}$g/cell~\cite{loffler12}, or $1.2\times10^{-4}$g/L.

Therefore, the particulate $\mathrm{COD_{X}}$ can be calculated by

\begin{align*}
\mathrm{COD_X} & \mathrm{= 2 \ g COD/(g \ biomass) \times 1.2\times10^{-4} (g \ biomass)/L} \\
& \mathrm{=2.4\times10^{-4} \ gCOD/L}.
\end{align*}

The COD of the substrate per COD of biomass ($k_{m,\mathrm{ch}}$) was then calculated using released chlor ions up to $35\mu$M/Ld, which corresponds to COD consumed by biomass of

\begin{align}
& \frac{35}{2.4\times10^{-4}} \frac{\mu\mathrm{mol[Cl^-]/Ld}}{\mathrm{gCOD_{X}/L}} = 0.14\mathrm{mol[Cl^-]/gCOD_{X}d}. \nonumber
\end{align}

It should be noted that this result does not include monochlorophenol, which have to be deduced. For monochlorophenol, the COD must be converted from the COD of Cl$^{-}$. By considering the stoichiometry of its oxygenation,

\begin{align}
\mathrm{C_{6}H_{5}OCl} + 6.5\mathrm{O_{2}} \rightarrow 6\mathrm{CO_2} + 2\mathrm{H_{2}O} + 2\mathrm{H^{+}} + 2\mathrm{Cl^{-}},
\end{align}
\noindent the COD of monochlorophenol can be calculated as $6.5 \times 32 = 208\mathrm{gCOD/mol}$. From this, the maximum specific uptake rate for monochlorophenol degraders can be estimated as
$k_{m,\mathrm{ch}} = 0.14 \times 208 = 29.12\mathrm{gCOD_{S}/gCOD_{X}d}$, where $\mathrm{COD_{S}}$ is the substrate COD.

For the half-saturation constant, $K_{S,\mathrm{ch}}$, no data was found detailing empirical studies for monochlorophenol degradation. However, from a recent study on the anaerobic microbial dechlorination of pentachlorophenol, some values were cited for dechlorinating bacteria using co-substrates of the 5-CP with other carbon sources~\cite{li10}. A $K_{S}$ value of $0.25\mu\mathrm{M}$, which converts to $0.053\mathrm{kgCOD/m^{3}}$ following the oxidation of monochlorophenol, was selected as a rough estimate for the chlorophenol degrader in this model.

The yield constant, $Y_{\mathrm{ch}}$ was calculated using information about the growth yield of the \textit{Dehalococcoides mccartyi} strain CBDB1 on 2,3-dichlorophenol (DCP), in which $7.6\times10^{13}$ cells per mole of Cl$^{-}$ were produced~\cite{adrian07}. Since yield is the ratio of $\mathrm{COD_{X}}$ to $\mathrm{COD_{S}}$, it can be calculated by determining these components from their stoichiometry;\\

\noindent $\bm{\mathrm{COD_{X}}}$: For $7.6\times10^{13}$ cells, the dry weight is calculated as $7.6\times10^{13}\times1.2\times10^{-14}\mathrm{g/cell} = 0.912\mathrm{g}$. Knowing that $2 \mathrm{gCOD/g biomass}$, stated earlier, then the $\mathrm{COD_{X} = 1.824\mathrm{g}}$.\\

\noindent $\bm{\mathrm{COD_{S}}}$: One mole of Cl$^{-}$ released should be converted to COD that DCP consumed. It is known that: DCP$\rightarrow2\mathrm{Cl}^{-}$. That means one mole of Cl$^{-}$ comes from $0.5$ mole of DCP.
The oxidation equation of DCP is given by

\begin{align}
\mathrm{C_{6}H_{4}OCl_{2}} + 6\mathrm{O_{2}} \rightarrow 6\mathrm{CO_{2}} +  \mathrm{H_{2}O} + 2\mathrm{H^{+}} + 2\mathrm{Cl^{-}}.
\end{align}

Therefore, $1$ mole of DCP consumes $6$ mole of oxygen, giving $6 \times 32 = 192\mathrm{gCOD}$, which corresponds to $0.5$ mole of DCP using $96\mathrm{gCOD}$.

Subsequently, the yield of the chlorophenol degrader can now be calculated as

\begin{align}
& Y_{\mathrm{ch}} = 1.824\mathrm{gCOD} / 96\mathrm{gCOD} = 0.019\mathrm{gCOD_{X}/gCOD_{S}}. \nonumber
\end{align}

The assumed half-saturation constant describing the affinity of the dechlorinator for hydrogen, $K_{S,\mathrm{H_{2}},c}$ was arbitrarily set to $1\times10^{-6}$ in the absence of supportable empirical data. The value was based on the known constant for the methanogen, and chosen to be an order of magnitude lower, accordingly. 

\paragraph{Parameters for phenol}

There is very limited data available on the kinetics of phenol degradation in anaerobic environments. The parameters selected for the phenol degrader here are approximate values based on available literature and deduction from analogous substrate reactions.

For example, the reaction of phenol is similar to that for propionate, but consumes twice as much oxygen. The oxidation equations for the two chemicals are written as

\begin{align}
\mathrm{Propionate}: \ \mathrm{C_{3}H_{6}O_{2}} + 3.5\mathrm{O_{2}} \rightarrow 3\mathrm{CO_{2}} + 3\mathrm{H_{2}O} \nonumber \\
\mathrm{Phenol}: \ \mathrm{C_{6}H_{5}OH} + 7\mathrm{O_{2}} \rightarrow 6\mathrm{CO_{2}} + 3\mathrm{H_{2}O}. \nonumber
\end{align}

Hence, the activity of phenol is assumed to be double the activity of propionate. From the ADM1~\cite{batstone02}, the maximum specific activity for propionate ($k_{m,\mathrm{pro}}$) is $13\mathrm{gCOD_{S}/gCOD_{X}}$, so $k_{m,\mathrm{ph}}$ is assumed to be $26\mathrm{gCOD_{S}/gCOD_{X}}$.
 
The half-saturation constant for phenol was taken from a study by Eismann \textit{et al.}~\cite{eismann97} using a modified Haldane kinetic model from Edwards~\cite{edwards70}, which considers the inhibitory effect of high substrate loading, for data fitting.  At 35$^\circ$C, the $K_{S,\mathrm{ph}}$ determined by the authors was $127 \mathrm{mgPhenol/l}$, which converts to a value of $0.302\mathrm{kgCOD/m^{3}}$ used in this model.
 
In the absence of compelling literature data, the yield parameter for the phenol degraders, $Y_{\mathrm{ph}}$, was taken to be equivalent to that for another acetogenic bacteria, the propionate degrader described in previous work~\cite{xu11} and having a value of $0.04\mathrm{gCOD_{S}/gCOD_{X}}$.

\section{Numerical Methods}\label{appendix:numerics}

Although it is possible to simplify the system and obtain analytical expressions for characteristic polynomials for all of the steady-states solved in~\ref{appendix: full_model}, it is impossible to gather the necessary information from these to determine the regions where each steady-state is meaningful and/or stable. Therefore, by considering sets of parameters ($D$, $S_{\mathrm{ch,in}}$, $S_{\mathrm{ph,in}}$ and $S_{\mathrm{H_2,in}}$) in turn on each steady-state, and through the use of Matlab, the complex polynomials for each steady-state in turn can be solved to determine which are meaningful and stable. Repeating this method with numerous sets of parameters leads to results showing where each steady-state is stable. 

In addition, some cases (SS3 or SS4) do not allow straightforward expressions for the steady-state (quadratic and cubic equations for $s_0$ are obtained, respectively). In these situations, the above procedure is followed for all possible values of $s_0$ at these steady-states.

\section{Stability analysis for the three-tier chlorophenol model with substrate addition}\label{appendix: full_model}

Here, the expressions used for each steady-states in turn when hydrogen and phenol addition are included in the model are given. To obtain similar expressions when hydrogen and/or phenol addition is not present in the model, simply set $u_h=0$ and/or $u_p=0$, respectively. For the two-tier phenol model, consider $s_0 = x_0 = 0$ and solve accordingly. 

\subsection*{SS1 ($x_{0} = x_{1} = x_{2} = 0)$}

Equations~\eqref{eqd1}, ~\eqref{eqd3} and~\eqref{eqd5} result in $s_{0} = u_{f}$, $s_{1} = u_{g}$ and $s_{2} = u_{h}$, such that the steady-state is always meaningful.

The Jacobian matrix, $J_{\rm SS1}$, can be written as

\[\left[ {\begin{array}{@{}cccccc@{}}
-\alpha+\mu_{0}-k_{A} & 0 & 0 & 0 & 0 & 0 \\
0 & -\alpha +\mu_1- k_{B} & 0 & 0 & 0 & 0 \\
0 & 0 & -\alpha+\mu_{2}-k_{C} & 0 & 0 & 0 \\
-\mu_{0} & 0 & 0 & -\alpha & 0 & 0 \\
\omega_{0}\mu_{0} & -\mu_1 & 0 & 0 & -\alpha & 0 \\
-\omega_{2}\mu_{0} & \omega_1\mu_1 & -\mu_{2} & 0 & 0 & -\alpha \end{array}} \right].\] 

Its six eigenvalues are $\lambda_{1} = -\alpha+\mu_{0}-k_{A}$, $\lambda_{2} = -\alpha+\mu_1-k_{B}$, $\lambda_{3} = -\alpha+\mu_{2}-k_{C}$, and $\lambda_{4} = \lambda_{5} = \lambda_{6} = -\alpha$. The condition for stability is $\lambda_{1,2,3} < 0$. By substituting $s_0$, $s_1$ and $s_2$ into each of $\mu_0$, $\mu_1$ and $\mu_2$, stability is assured when 

\begin{align}
&\frac{u_{f}}{1+u_{f}}\frac{u_{h}}{K_{P}+u_{h}}-k_{A} < \alpha; \ \frac{\phi_1u_{g}}{1+u_{g}}\frac{1}{1+K_Iu_{h}}-k_{B} < \alpha; \nonumber \\
&\frac{\phi_{2}u_{h}}{1+u_{h}} - k_{C} < \alpha. \nonumber
\end{align}

\subsection*{SS2 ($x_{0} = 0, x_{1} = 0, x_{2} > 0)$}

Equations~\eqref{eqd1} and~\eqref{eqd3} give $s_{0} = u_{f}$ and $s_{1} = u_g$. Solving the remaining dimensionless equations gives

\begin{align}
s_2 &= -\frac{\alpha+k_C}{\alpha+k_C-\phi_2} \label{eq:b1}\\
x_2 &= \frac{\alpha\left(u_h-s_2\right)}{\alpha+k_C} \label{eq:b2}
\end{align}

Here, the Jacobian matrix, $J_{\rm SS2}$, then becomes

\[\left[ {\begin{array}{@{}cccccc@{}}
-\alpha+\mu_{0}-k_{A} & 0 & 0 & 0 & 0 & 0 \\
0 & -\alpha +\mu_1- k_{B} & 0 & 0 & 0 & 0 \\
0 & 0 & 0 & 0 & 0 & Ix_{2} \\
-\mu_{0} & 0 & 0 & -\alpha & 0 & 0 \\
\omega_{0}\mu_{0} & -\mu_1 & 0 & 0 & -\alpha & 0 \\
-\omega_{2}\mu_{0} & \omega_1\mu_1 & -(\alpha+k_{C}) & 0 & 0 & -\alpha-Ix_{2} \end{array}} \right].\] 

Its eigenvalues are $\lambda_{1} = -\alpha+\mu_{0}-k_{A}$, $\lambda_{2} = -\alpha+\mu_1-k_{B}$, $\lambda_{3} = \lambda_{4} = -\alpha$, $\lambda_{5,6} = \dfrac{-Ix_{2}+\alpha}{2}\pm\left[\dfrac{(Ix_{2}+\alpha)^2}{4}-\alpha Ix_{2}-k_{C}Ix_{2} \right]^{1/2}$. For stability, it is simply required that $\lambda_{1,2} < 0$, as for all meaningful values of $\alpha$, $I$, $x_2$ and $k_c$, there is always $\lambda_{3,4,5,6}<0$.

\subsection*{SS3: ${x_{0} > 0, x_{1} = 0, x_{2} = 0}$}

Solving Eqs~\eqref{eqd1}, \eqref{eqd2}, \eqref{eqd3} and~\eqref{eqd5} gives

 \begin{align}
  s_0^2\omega_2&(1-\alpha-k_A)+s_0[\omega_2u_f(\alpha+A-1)+u_h(1-\alpha-k_A) \nonumber \\
  & -(\alpha+A)(K_{P}+\omega_2)]-(\alpha+k_A)(K_{P}-\omega_2u_f+u_h)=0 \label{eq:b3} \\
  s_2 &= u_h-\omega_2\left(u_f-s_0\right)  \label{eq:b4} \\
  x_0 &= \frac{\alpha\left(u_f-s_0\right)}{\alpha+k_A}  \label{eq:b5} \\
  s_1 &= u_g+\frac{\omega_0\left(\alpha+k_A\right)x_0}{\alpha}.
\end{align}
As mentioned previously in~\ref{appendix:numerics}, when numerically checking whether this steady-state is meaningful, each possible value for $s_0$ is considered in turn.

The Jacobian matrix, $J_{\rm SS3}$, then becomes

\[\left[ \scalemath{0.8} {\begin{array}{@{}cccccc@{}}
0 & 0 & 0 & Ex_{0} & 0 & Fx_{0} \\
0 & -\alpha-+\mu_1-k_{B} & 0 & 0 & 0 & 0 \\
0 & 0 & -\alpha+\mu_{2}-k_{C} & 0 & 0 & 0 \\
-(\alpha+k_{A}) & 0 & 0 & -\alpha - Ex_{0} & 0 & -Fx_{0} \\
\omega_{0}(\alpha+k_{A}) & -\mu_{1} & 0 & \omega_{0}Ex_{0} & -\alpha & \omega_{0}Fx_{0} \\
-\omega_{2}(\alpha+k_{A}) & \omega_{1}\mu_{1} & -\mu_{2} & -\omega_{2}Ex_{0} & 0 & -\alpha-\omega_{2}Fx_{0} \end{array}} \right].\] 

Its known eigenvalues are $\lambda_1=-\alpha+\mu_1-k_B$, $\lambda_2=-\alpha+\mu_2-k_C$, $\lambda_3=-\alpha$. The other three eigenvalues are given by the characteristic polynomial

\begin{equation}
  \lambda^3+f_2\lambda^2+f_1\lambda+f_0=0
\label{eq:SS3_poly}
\end{equation}
where

\begin{align}
  f_2 &= 2\alpha + Ex_0 + F\omega_2x_0\\
  f_1 &= \alpha^2 + k_AEx_0 + 2E\alpha x_0 + k_AF\omega_2x_0 + 2F\alpha\omega_2x_0\\
  f_0 &= E\alpha^2x_0 + F\alpha^2\omega_2x_0 + k_AE\alpha x_0 + k_AF\alpha\omega_2x_0.
\end{align}
For stability it is required that $\lambda_{1,2} <0$ and all three roots from Eq.~\eqref{eq:SS3_poly} have negative real parts. These are checked following the procedure described in~\ref{appendix:numerics}.

\subsection*{SS4 ($x_{0} > 0, x_{1} > 0, x_{2} = 0)$}
By solving Eqs.~\eqref{eqd2} to~\eqref{eqd5} for $x_0$, $x_1$, $s_1$ and $s_2$, and substituting these into equation~\eqref{eqd1}, the following cubic expression for $s_0$ is obtained

\begin{equation}
  p_3s_0^3+p_2s_0^2+p_1s_0+p_0=0 \label{eq:b11}
\end{equation}
where

\begin{align}
  \nonumber p_3 &= \left(k_A+\alpha\right)\left(\left(k_B+\alpha\right) \left(k_A+\alpha-K_IK_{P}\left(k_A+\alpha\right)\right) \right. \\ \nonumber & - \left. \phi_1\left(k_A+\alpha\right)\right)\left(u_h-\omega_1u_g+\alpha u_f\left(\omega_2-\omega_0\omega_1\right)\right) \\ \nonumber & -K_{P}\left(k_A+\alpha\right)\left(\left(k_B+\alpha\right)\left(k_A+\alpha-K_IK_{P}\left(k_A+\alpha\right)\right) \right. \\ \nonumber & \left. - \phi_1\left(k_A+\alpha\right)\right) -\omega_1\left(k_A+\alpha\right)\left(k_B+\alpha\right) \\ & \left(k_A+\alpha-K_IK_{P}\left(k_A+\alpha\right)\right)
      \end{align}
        \begin{align}
  \nonumber p_2 &= \left(\left(\left(k_B+\alpha\right)\left(k_A+\alpha-K_IK_{P}\left(k_A+\alpha\right)\right)-\phi_1\left(k_A+\alpha\right)\right) \right. \\ \nonumber & \left. \left(k_A+\alpha-1\right) +\left(k_A+\alpha\right)\left(\left(k_B +\alpha\right)\right.\right. \\ \nonumber & \left.\left. \left(k_A+\alpha-K_IK_{P}\left(k_A+\alpha\right)-1\right) \phi_1\left(k_A+\alpha-1\right)\right)\right) \\ \nonumber & \left(u_h-\omega_1u_g+\alpha u_f\left(\omega_2-\omega_0\omega_1\right)\right) -K_{P}\left(k_A+\alpha\right)  \\ \nonumber &  \left(\left(k_B+\alpha\right)\left(k_A+\alpha-K_IK_{P}\left(k_A+\alpha\right)\right) - \phi_1\left(k_A+\alpha\right)\right) \\ \nonumber & -K_{P}\left(k_A+\alpha\right)\left(\left(k_B+\alpha\right)\left(k_A+\alpha-K_IK_{P}\left(k_A+\alpha\right)-1\right) \right. \\ \nonumber & - \left. \phi_1\left(k_A+\alpha-1\right)\right)-\alpha\left(k_A+\alpha\right)\left(\omega_2-\omega_0\omega_1\right)\left(\left(k_B+\alpha\right)\right. \\ \nonumber & \left.\left(k_A+\alpha-K_IK_{P}\left(k_A+\alpha\right)\right)-\phi_1\left(k_A+\alpha\right)\right)  \\ \nonumber & -\omega_1\left(k_A+\alpha\right)\left(k_B+\alpha\right)\left(k_A+\alpha -K_IK_{P}\left(k_A+\alpha\right)-1\right) \\ &  -\omega_1\left(k_B+\alpha\right)\left(k_A+\alpha-1\right)\left(k_A+\alpha-K_IK_{P}\left(k_A+\alpha\right)\right)\\
  \nonumber p_1 &= \left(\left(k_B+\alpha\right)\left(k_A+\alpha-K_IK_{P}\left(k_A+\alpha\right)-1\right) \right. \\ \nonumber & \left. -\phi_1\left(k_A+\alpha-1\right)\right)\left(k_A+\alpha-1\right)\left(u_h-\omega_1u_g+\alpha u_f\left(\omega_2-\omega_0\omega_1\right)\right) \\ \nonumber &  -K_{P}\left(k_A+\alpha\right)\left(\left(k_B+\alpha\right)\left(k_A+\alpha-K_IK_{P}\left(k_A+\alpha\right)-1\right) \right. \\ \nonumber & - \left. \phi_1\left(k_A+\alpha-1\right)\right)-\alpha\left(\omega_2-\omega_0\omega_1\right)\left(\left(\left(k_B+\alpha\right) \right.\right. \\ \nonumber & \left.\left. \left(k_A+\alpha-K_IK_{P}\left(k_A+\alpha\right)\right)  \phi_1\left(k_A+\alpha\right)\right)\left(k_A+\alpha-1\right) \right. \\ \nonumber & \left. +\left(k_A+ \alpha\right)\left(\left(k_B+\alpha\right)\left(k_A+\alpha-K_IK_{P}\left(k_A+\alpha\right)-1\right) \right.\right. \\ \nonumber & - \left.\left. \phi_1\left(k_A+\alpha-1\right)\right)\right)  -\omega_1\left(k_B+\alpha\right) \\ & \left(k_A+\alpha-1\right)\left(k_A+\alpha-K_IK_{P}\left(k_A+\alpha\right)-1\right) \\
  \nonumber p_0 &= \alpha\left(\omega_2-\omega_0\omega_1\right)\left(\left(k_B+\alpha\right)\left(k_A+\alpha-K_IK_{P}\left(k_A+\alpha\right)-1\right) \right. \\ & - \left. \phi_1\left(k_A+\alpha-1\right)\right)\left(k_A+\alpha-1\right).
\end{align}

Once again, when numerically checking whether this steady-state is meaningful, each possible value for $s_0$ in turn is checked. The remaining expressions for this steady-state are given by

  \begin{align}
  s_2 &= \frac{-K_{P}\left(\alpha+k_A\right)\left(1+s_0\right)}{\left(1+s_0\right)\left(\alpha+k_A\right)-s_0}\\
  s_1 &=\frac{-\left(1+K_Is_2\right)\left(\alpha+k_B\right)}{\left(1+K_Is_2\right)\left(\alpha+k_B\right)-\phi_1}\\
  x_0 &=\frac{\alpha\omega_1\left(u_g-s_1\right)+\alpha\left(u_h-s_2\right)}{\left(\alpha+k_A\right)\left(\omega_2-\omega_0\omega_1\right)} \label{eq:b18}\\
  x_1 &=\frac{\omega_2\left(\alpha+k_A\right)}{\omega_1\left(\alpha+k_B\right)}x_0-\frac{\alpha}{\omega_1\left(\alpha+k_B\right)}\left(u_h-s_2\right).
\end{align}

The Jacobian matrix, $J_{\rm SS4}$, becomes

\[\left[ \scalemath{0.74} {\begin{array}{@{}cccccc@{}}
0 & 0 & 0 & Ex_{0} & 0 & Fx_{0} \\
0 & 0 & 0 & 0 & Gx_{1} & Hx_{1} \\
0 & 0 & -\alpha+\mu_{2}-k_{C} & 0 & 0 & 0 \\
-(\alpha+k_{A}) & 0 & 0 & -\alpha-Ex_{0} & 0 & -Fx_{0} \\
\omega_{0}(\alpha+k_{A}) & -(\alpha+k_{B})  & 0 & \omega_{0}Ex_{0} & -\alpha-Gx_{1} & \omega_{0}Fx_{0}-Hx_{1} \\
-\omega_{2}(\alpha+k_{A}) & \omega_{1}(\alpha+k_{B})  & -\mu_{2} & -\omega_{2}Ex_{0} & \omega_{1}Gx_{1} & -\alpha-\omega_{2}Fx_{0}+\omega_{1}Hx_{1} \end{array}} \right].\] 

Its known eigenvalue is $\lambda_1=-\alpha+\mu_2-k_C$ and the characteristic polynomial for the remaining five eigenvalues is given by

\begin{equation}
  \lambda^5+f_4\lambda^4+f_3\lambda^3+f_2\lambda^2+f_1\lambda+f_0=0
\label{eq:SS4_poly}
\end{equation}
where

\begin{align}
  f_4 &= 3\alpha+Ex_0+Gx_1+F\omega_2x_0-H\omega_1x_1 \\
  f_3 &= 3\alpha^2+k_AEx_0+k_BGx_1+3E\alpha x_0+3G\alpha x_1 \nonumber \\ & +k_AF\omega_2x_0-k_BH\omega_1x_1+EGx_0x_1+3F\alpha\omega_2x_0 \nonumber \\ & -3H\alpha\omega_1x_1+FG\omega_2x_0x_1-EH\omega_1x_0x_1\nonumber \\ & -FG\omega_0\omega_1x_0x_1 \\
  f_2 &= \alpha^3+3E\alpha^2x_0+3G\alpha^2x_1+3F\alpha^2\omega_2x_0- 3H\alpha^2\omega_1x_1 \nonumber \\ & +2k_AE\alpha x_0+2k_BG\alpha x_1+k_AEGx_0x_1+k_BEGx_0x_1\nonumber \\ &+2k_AF\alpha\omega_2x_0-2k_BH\alpha\omega_1x_1+3EG\alpha x_0x_1\nonumber \\ &+k_AFG\omega_2x_0x_1 -k_AEH\omega_1x_0x_1+k_BFG\omega_2x_0x_1\nonumber \\ & -k_BEH\omega_1x_0x_1+3FG\alpha\omega_2x_0x_1-3EH\alpha\omega_1x_0x_1\nonumber \\ &-3FG\alpha\omega_0\omega_1x_0x_1-k_AFG\omega_0\omega_1x_0x_1\nonumber \\ &-k_BFG\omega_0\omega_1x_0x_1\\
  f_1 &= E\alpha^3x_0+G\alpha^3x_1+k_AE\alpha^2x_0+k_BG\alpha^2x_1+F\alpha^3\omega_2x_0\nonumber \\ &-H\alpha^3\omega_1x_1+k_AF\alpha^2\omega_2x_0-k_BH\alpha^2\omega_1x_1+3k_EG\alpha^2x_0x_1\nonumber \\ &+3FG\alpha^2\omega_2x_0x_1-3EH\alpha^2\omega_1x_0x_1+k_Ak_BEGx_0x_1\nonumber \\ &+2k_AEG\alpha x_0x_1+2k_BEG\alpha x_0x_1-3FG\alpha^2\omega_0\omega_1x_0x_1\nonumber \\ &+k_Ak_BFG\omega_2x_0x_1-k_Ak_BEH\omega_1x_0x_1+2k_AFG\alpha\omega_2x_0x_1\nonumber \\ &-2k_AEH\alpha\omega_1x_0x_1+2k_BFG\alpha\omega_2x_0x_1-2k_BEH\alpha\omega_1x_0x_1\nonumber \\ &-k_Ak_BFG\omega_0\omega_1x_0x_1- 2k_AFG\alpha\omega_0\omega_1x_0x_1\nonumber \\ &-2k_BFG\alpha\omega_0\omega_1x_0x_1 \\
  f_0 &= EG\alpha^3x_0x_1+FG\alpha^3\omega_2x_0x_1-EH\alpha^3\omega_1x_0x_1\nonumber \\ &+k_AEG\alpha^2x_0x_1+k_BEG\alpha^2x_0x_1+k_AFG\alpha^2\omega_2x_0x_1\nonumber \\ &-k_AEH\alpha^2\omega_1x_0x_1+k_BFG\alpha^2\omega_2x_0x_1-k_BEH\alpha^2\omega_1x_0x_1\nonumber \\ &-FG\alpha^3\omega_0\omega_1x_0x_1+k_Ak_BEG\alpha x_0x_1\nonumber \\ &+k_Ak_BFG\alpha\omega_2x_0x_1-k_Ak_BEH\alpha\omega_1x_0x_1\nonumber \\ &-k_AFG\alpha^2\omega_0\omega_1x_0x_1-k_BFG\alpha^2\omega_0\omega_1x_0x_1\nonumber \\ &-k_Ak_BFG\alpha\omega_0\omega_1x_0x_1.
\end{align}
For stability it is required that $\lambda_{1} <0$ and all five roots from equation~\eqref{eq:SS4_poly} have negative real parts.

\subsection*{SS5 ($x_{0} > 0, x_{1} = 0, x_{2} > 0$)}

By solving the dimensionless equations~\eqref{eqd1}, \eqref{eqd2}, \eqref{eqd3}, \eqref{eqd5} and~\eqref{eqd6}, the steady-state is given by

\begin{align}
  s_2 &= \frac{-\alpha-k_C}{\alpha+k_C-\phi_2} \\
  s_0 &= \frac{-\left(K_{P}+s_2\right)\left(\alpha+k_A\right)}{\left(K_{P}+s_2\right)\left(\alpha+k_A\right)-s_2} \\
  x_0 &=\frac{\alpha\left(u_f-s_0\right)}{\alpha+k_A} \\
  x_2 &=\frac{\alpha\left(u_h-s_2\right)-\omega_2\left(\alpha+k_A\right)x_0}{\alpha+k_C}  \label{eq:b29} \\
  s_1 &=u_g+\frac{\omega_0\left(\alpha+k_A\right)x_0}{\alpha}.
\end{align}

The Jacobian matrix,$J_{\rm SS5}$, becomes

\[\left[\scalemath{0.83} {\begin{array}{@{}cccccc@{}}
  0 & 0 & 0 & Ex_0 & 0 & Fx_0 \\
  0 & -\alpha+\mu_1-k_B & 0 & 0 & 0 & 0 \\
  0 & 0 & 0 & 0 & 0 & Ix_2 \\
  \alpha+k_A & 0 & 0 & -\alpha-Ex_0 & 0 & -Fx_0 \\
  \omega_0\left(\alpha+k_A\right) & -\mu_1 & 0 & E\omega_0x_0 & -\alpha & F\omega_0x_0 \\
  -\omega_2\left(\alpha+k_A\right) & \omega_1\mu_1 & -\left(\alpha+k_C\right) & -E\omega_2x_0 & 0 & -\alpha-F\omega_2x_0-Ix_2
\end{array}} \right].\]

Its known eigenvalues are $\lambda_1=-\alpha+\mu_1-k_B$ and $\lambda_2=-\alpha$. The remaining four are given by the characteristic polynomial

\begin{equation}
  \lambda^4+f_3\lambda^3+f_2\lambda^2+f_1\lambda+f_0=0
\label{eq:SS5_poly}
\end{equation}
where

\begin{align}
  f_3 &= 2\alpha+Ex_0+Ix_2+F\omega_2x_0 \\
  f_2 &= \alpha^2+k_AEx_0+CIx_2+2E\alpha x_0+2I\alpha x_2+k_AF\omega_2x_0+ \nonumber \\ &EIx_0x_2+2F\alpha\omega_2x_0 \\
  f_1 &= E\alpha^2x_0+I\alpha^2x_2+F\alpha^2\omega_2x_0+k_AE\alpha x_0+ k_CI\alpha x_2\nonumber \\ &+k_AEIx_0x_2+k_CEIx_0x_2+k_AF\alpha\omega_2x_0+2EI\alpha x_0x_2 \\
  f_0 &= EI\alpha^2x_0x_2+k_ACEIx_0x_2+k_AEI\alpha x_0x_2+k_CEI\alpha x_0x_2.
\end{align}
For stability it is required that $\lambda_{1,2} <0$ and all four roots from equation~\eqref{eq:SS5_poly} have negative real parts.

\subsection*{SS6 $(x_{0} > 0, x_{1} > 0, x_{2} > 0)$}
Solving all six dimensionless equations~\eqref{eqd2} to~\eqref{eqd5}, leads to the expressions for this steady-state

\begin{align}
  s_2 &= \frac{-\alpha-k_C}{\alpha+k_C-\phi_2} \\
  s_1 &= \frac{-\left(1+K_Is_2\right)\left(\alpha+k_B\right)}{\left(1+K_Is_2\right)\left(\alpha+k_B\right)-\phi_1} \\
  s_0 &= \frac{-\left(K_{P}+s_2\right)\left(\alpha+k_A\right)}{\left(K_{P}+s_2\right)\left(\alpha+k_A\right)-\phi_1} \\
  x_0 &= \frac{\alpha\left(u_f-s_0\right)}{\left(\alpha+k_A\right)} \\
  x_1 &= \frac{\alpha\left(u_g-s_1\right)+\omega_0\left(\alpha+k_A\right)x_0}{\alpha+k_B} \\
  x_2 &= \frac{\alpha\left(u_h-s_2\right)-\omega_2\left(\alpha+k_A\right)x_0+\omega_1\left(\alpha+k_B\right)x_1}{\alpha+k_C}.
\end{align}

The Jacobian matrix, $J_{\rm SS6}$, becomes

\begin{equation*}
 \left[\scalemath{0.74} {\begin{array}{@{}cccccc@{}}
0 & 0 & 0 & Ex_{0} & 0 & Fx_{0} \\
0 & 0 & 0 & 0 & Gx_{1} & Hx_{1} \\
0 & 0 & 0 & 0 & 0 & Ix_{2} \\
-(\alpha+k_{A}) & 0 & 0 & -\alpha-Ex_{0} & 0 & -Fx_{0} \\
\omega_{0}(\alpha+k_{A}) & -(\alpha+k_{B})  & 0 & \omega_{0}Ex_{0} & -\alpha-Gx_{1} & \omega_{0}Fx_{0}-Hx_{1} \\
-\omega_{2}(\alpha+k_{A}) & \omega_{1}(\alpha+k_{B})  & -(\alpha+k_{C}) & -\omega_{2}Ex_{0} & \omega_{1}Gx_{1} & -\alpha-\omega_{2}Fx_{0}+\omega_{1}Hx_{1}-Ix_{2} \end{array}} \right].
\end{equation*}

Here, all six eigenvalues are given by the characteristic polynomial

\begin{equation}
  \lambda^6+f_5\lambda^5+f_4\lambda^4+f_3\lambda^3+f_2\lambda^2+f_1\lambda+f_0=0
\label{eq:SS6_poly}
\end{equation}
where

\begin{align}
  f_5 &= 3\alpha+Ex_0+Gx_1+Ix_2+F\omega_2x_0-H\omega_1x_1 \\
  f_4 &= 3\alpha^2+k_AEx_0+k_BGx_1+k_CIx_2+3E\alpha x_0+3G\alpha x_1\nonumber \\ &+ 3I\alpha x_2+k_AF\omega_2x_0-k_BH\omega_1x_1+EGx_0x_1+EIx_0x_2\nonumber \\ &+GIx_1x_2+3F\alpha\omega_2x_0-3H\alpha\omega_1x_1+FG\omega_2x_0x_1\nonumber \\ &-EH\omega_1x_0x_1-FG\omega_0\omega_1x_0x_1 \\
  f_3 &= \alpha^3+3E\alpha^2x_0+3G\alpha^2x_1+3I\alpha^2x_2+3F\alpha^2\omega_2x_0\nonumber \\ &-3H\alpha^2\omega_1x_1+2k_AE\alpha x_0+2k_BG\alpha x_1+2k_CI\alpha x_2\nonumber \\ &+k_AEGx_0x_1+k_BEGx_0x_1+k_AEIx_0x_2+k_CEIx_0x_2\nonumber \\ &+k_BGIx_1x_2+k_CGIx_1x_2+2k_AF\alpha\omega_2x_0-2k_BH\alpha\omega_1x_1\nonumber \\ &+3EG\alpha x_0x_1+3EI\alpha x_0x_2+3GI\alpha x_1x_2+k_AFG\omega_2x_0x_1\nonumber \\ &-k_AEH\omega_1x_0x_1+k_BFG\omega_2x_0x_1-k_BEH\omega_1x_0x_1\nonumber \\ &+EGIx_0x_1x_2+3FG\alpha\omega_2x_0x_1-3EH\alpha\omega_1x_0x_1\nonumber \\ &-3FG\alpha\omega_0\omega_1x_0x_1-k_AFG\omega_0\omega_1x_0x_1\nonumber \\ &-k_BFG\omega_0\omega_1x_0x_1 \\
  f_2 &= E\alpha^3x_0+G\alpha^3x_1+I\alpha^3x_2+k_AE\alpha^2x_0+k_BG\alpha^2x_1\nonumber \\ &+k_CI\alpha^2x_2+F\alpha^3\omega_2x_0-H\alpha^3\omega_1x_1+k_AF\alpha^2\omega_2x_0\nonumber \\ &-k_BH\alpha^2\omega_1x_1+3EG\alpha^2x_0x_1+3EI\alpha^2x_0x_2 +3GI\alpha^2x_1x_2\nonumber \\ & +3FG\alpha^2\omega_2x_0x_1-3EH\alpha^2\omega_1x_0x_1+k_Ak_BEGx_0x_1\nonumber \\ &+k_Ak_CEIx_0x_2+k_Bk_CGIx_1x_2+2k_AEG\alpha x_0x_1\nonumber \\ &+2k_BEG\alpha x_0x_1+2k_AEI\alpha x_0x_2+2k_CEI\alpha x_0x_2\nonumber \\ &+2k_BGI\alpha x_1x_2+2k_CGI\alpha x_1x_2-3FG\alpha^2\omega_0\omega_1x_0x_1\nonumber \\ &+k_Ak_BFG\omega_2x_0x_1-k_Ak_BEH\omega_1x_0x_1+k_AEGIx_0x_1x_2\nonumber \\ &+k_BEGIx_0x_1x_2+k_CEGIx_0x_1x_2+2k_AFG\alpha\omega_2x_0x_1\nonumber \\ &-2k_AEH\alpha\omega_1x_0x_1+2k_BFG\alpha\omega_2x_0x_1-2k_BEH\alpha\omega_1x_0x_1\nonumber \\ &+3EGI\alpha x_0x_1x_2-k_Ak_BFG\omega_0\omega_1x_0x_1-2k_AFG\alpha\omega_0\omega_1x_0x_1\nonumber \\ &-2k_BFG\alpha\omega_0\omega_1x_0x_1 \\
  f_1 &= E\alpha^3x_0+G\alpha^3x_1+k_AE\alpha^2x_0+k_BG\alpha^2x_1+F\alpha^3\omega_2x_0\nonumber \\ &-H\alpha^3\omega_1x_1+k_AF\alpha^2\omega_2x_0-k_BH\alpha^2\omega_1x_1+3EG\alpha^2x_0x_1\nonumber \\ &+3FG\alpha^2\omega_2x_0x_1-3EH\alpha^2\omega_1x_0x_1+k_Ak_BEGx_0x_1\nonumber \\ &+2k_AEG\alpha x_0x_1+2k_BEG\alpha x_0x_1-3FG\alpha^2\omega_0\omega_1x_0x_1\nonumber \\ &+k_Ak_BFG\omega_2x_0x_1-k_Ak_BEH\omega_1x_0x_1+2k_AFG\alpha\omega_2x_0x_1\nonumber \\ &-2k_AEH\alpha\omega_1x_0x_1+2k_BFG\alpha\omega_2x_0x_1-2k_BEH\alpha\omega_1x_0x_1\nonumber \\ &-k_Ak_BFG\omega_0\omega_1x_0x_1- 2k_AFG\alpha\omega_0\omega_1x_0x_1\nonumber \\ &-2k_BFG\alpha\omega_0\omega_1x_0x_1 \\
  f_0 &= EGI\alpha^3x_0x_1x_2+k_AEGI\alpha^2x_0x_1x_2+k_BEGI\alpha^2x_0x_1x_2 \nonumber \\ &+k_CEGI\alpha^2x_0x_1x_2+k_Ak_Bk_CEGIx_0x_1x_2\nonumber \\ &+k_Ak_BEGI\alpha x_0x_1x_2+k_Ak_CEGI\alpha x_0x_1x_2\nonumber \\ &+k_Bk_CEGI\alpha x_0x_1x_2.
\end{align}
For stability it is required that all six roots from equation~\eqref{eq:SS6_poly} have negative real parts.

\subsection*{SS7 $(x_{0} = 0, x_{1} >0, x_{2} = 0)$}
Solving  equations~\eqref{eqd1},~\eqref{eqd3},~\eqref{eqd4} and~\eqref{eqd5} gives

\begin{align}
  & s_1^2\left[K_{I}u_h\left(\alpha+k_{B}\right) - \alpha - k_{B} + K_{I}\omega_1u_h\left(\alpha+k_{B}\right)\right]+\nonumber \\ &s_1\left[K_{I}u_h(\alpha+k_{B}) - \alpha - \phi_1 - K_{I}\omega_1\left(\alpha+k_{B}\right) - k_{B} \right. \nonumber \\ & +\left. K_{I}\omega_1u_h\left(\alpha+k_{B}\right)\right]-  K_{I}\omega_1\left(\alpha+k_{B}\right)=0 \\
  & s_2=\frac{\phi_1s_1}{\left(1+s_1\right)K_{I}\left(\alpha+k_{B}\right)}-\frac{1}{K_{I}} \label{eq:b49}\\
  & s_0=u_f \\
  & x_1=\frac{\alpha\left(u_g-s_1\right)}{\alpha+k_{B}}.
\end{align}

As mentioned previously in~\ref{appendix:numerics}, when numerically checking whether this steady-state is meaningful, each possible value for $s_0$ is considered in turn.

The Jacobian matrix, $J_{\rm SS7}$, becomes

\[\left[\scalemath{0.85} {\begin{array}{@{}cccccc@{}}
  -\alpha+\mu_0-k_{A} & 0  & 0 & 0 & 0 & 0 \\
  0 & 0 & 0 & 0 & Gx_1 & Hx_1 \\
  0 & 0 & -\alpha+\mu_2-k_C & 0 & 0 & 0 \\
  \mu_0 & 0 & 0 & -\alpha & 0 & 0 \\
  \omega_0\mu_0 & -\left(\alpha+k_{B}\right) & 0 & 0 & -\alpha-Gx_1 & -Hx_1 \\
  -\omega_2\mu_0 & \omega_1\left(\alpha+k_{B}\right) & -\mu_2 & 0 & G\omega_1x_1 & -\alpha+H\omega_1x_1 \end{array}} \right].\]

Its known eigenvalues are $\lambda_1=-\alpha+\mu_0-k_{A}$, $\lambda_2+-\alpha+\mu_2-k_{C}$ and $\lambda_3=-\alpha$. The other three eigenvalues are given by the characteristic polynomial

\begin{equation}
  \lambda^3+f_2\lambda^2+f_1\lambda+f_0=0
  \label{eq:poly_SS7}
\end{equation}
where

\begin{align}
  f_2 &= 2\alpha + Gx_1 - H\omega_1x_1 \\
  f_1 &= \alpha^2 + k_{B}Gx_1 + 2G\alpha x_1 - k_{B}H\omega_1x_1 - 2H\alpha\omega_1x_1 \\
  f_0 &= G\alpha^2x_1 - H\alpha^2\omega_1x_1 + k_{B}G\alpha x_1 - k_{B}H\alpha\omega_1x_1.
\end{align}

For stability we require $\lambda_{1,2,3}<0$ and all three roots from equation~\eqref{eq:poly_SS7} to have negative real parts. These are checked following the procedure described in~\ref{appendix:numerics}.

\subsection*{SS8 $(x_{0} = 0, x_{1} >0, x_{2} > 0)$}
By solving the dimensionless equations~\eqref{eqd1}, \eqref{eqd3}, \eqref{eqd4}, \eqref{eqd5} and \eqref{eqd6}, the steady-state is given by

\begin{align}
  s_2 &= \frac{-\alpha-k_{C}}{\alpha+k_{C}-\phi_2} \\
  s_1 &= \frac{-\left(1+K_{I}s_2\right)\left(\alpha+k_{B}\right)}{\left(1+K_{I}s_2\right)\left(\alpha+k_{B}\right)-\phi_1} \\
  x_1 &= \frac{\alpha\left(u_g-s_1\right)}{\alpha+k_{B}} \\
  x_2 &= \frac{\alpha\left(u_h-s_2\right)+\omega_1\left(\alpha+k_{B}\right)x_1}{\alpha+k_{C}}\\
  s_0 &= u_f
\end{align}
The Jacobian matrix, $J_{\rm SS8}$, becomes

\[\left[\scalemath{0.74} {\begin{array}{@{}cccccc@{}}
  -\alpha+\mu_0-k_A & 0 & 0 & 0 & 0 & 0 \\
  0 & 0 & 0 & 0 & Gx_1 & Hx_1 \\
  0 & 0 & 0 & 0 & 0 & Ix_2 \\
  \mu_0 & 0 & 0 & -\alpha & 0 & 0 \\
  \omega_0\mu_0 & -\left(\alpha+k_{B}\right) & 0 & 0 & -\alpha-Gx_1 & -Hx_1 \\
  -\omega_2\mu_0 & \omega_1\left(\alpha+k_{B}\right) & -\left(\alpha+k_C\right) & 0 & G\omega_1x_1 & -\alpha+H\omega_1x_1-Ix_2 \end{array}} \right].\]

Its known eigenvalues are $\lambda_1=-\alpha+\mu_0-k_A$ and $\lambda_2=-\alpha$. The remaining four eigenvalues are given by the characteristic polynomial 

\begin{equation}
  \lambda^4+f_3\lambda^3+f_2\lambda^2+f_1\lambda+f_0=0
  \label{eq:poly_SS8}
\end{equation}
where

\begin{align}
  f_3 &= 2\alpha+Gx_1+Ix_2-H\omega_1x_1 \\
  f_2 &= \alpha^2+k_{B}Gx_1+k_CIx_2+2G\alpha x_1+2I\alpha x_2-k_{B}H\omega_1x_1+\nonumber \\ &GIx_1x_2-2H\alpha\omega_1x_1 \\
  f_1 &= G\alpha^2x_1+I\alpha^2x_2-H\alpha^2\omega_1x_1+k_{B}G\alpha x_1+k_CI\alpha x_2\nonumber \\ &+k_{B}GIx_1x_2+k_CGIx_1x_2-k_{B}H\alpha\omega_1x_1\nonumber \\ &+2GI\alpha x_1x_2 \\
  f_0 &= GI\alpha^2x_1x_2+k_{B}k_CGIx_1x_2+k_{B}GI\alpha x_1x_2\nonumber \\ &+k_CGI\alpha x_1x_2.
\end{align}

For stability it is required that $\lambda_{1,2}<0$ and all four roots from equation~\eqref{eq:poly_SS8} to have negative real parts.

\section*{References}\label{Bibliography}
\bibliographystyle{elsarticle-num} 

\end{document}